\documentclass{aa}

\usepackage[varg]{txfonts}
\usepackage{booktabs}
\usepackage{threeparttable}
\usepackage{multirow}
\usepackage{siunitx}
\usepackage{pdflscape}
\usepackage{longtable}
\bibpunct{(}{)}{;}{a}{}{,} 

\begin{document}

\title{Systematic Differences in Spectroscopic Analysis of Red Giants
\thanks{Based on spectroscopic observations made with the Nordic Optical Telescope operated by NOTSA at the Observatorio del Roque de los Muchachos (La Palma, Spain) of the Instituto de Astrofísica de Canarias.}
}

\author{D. Slumstrup \inst{1}\fnmsep\thanks{ditte@phys.au.dk} 
		\and F. Grundahl \inst{1} 
        \and V. Silva Aguirre \inst{1} 
        \and K. Brogaard \inst{1}}

\institute{Stellar Astrophysics Centre (SAC). Department of Physics and Astronomy, Aarhus University, Ny Munkegade 120, DK-8000 Aarhus, Denmark} 

\date{Received 29 June 2018 / Accepted 11 December 2018 }

\abstract{}{A spectroscopic analysis of stellar spectra can be carried out using different approaches; different methods, line lists, atmospheric models, atomic parameters, solar abundances etc. The resulting atmospheric parameters from these choices can vary beyond the quoted uncertainties in the literature. Here we characterize these differences by systematically comparing some of the commonly adopted ingredients; line lists, equivalent width measurements and atomic parameters.}{High resolution and high signal-to-noise (S/N) spectroscopic data of one helium-core-burning red giant star in each of the three open clusters, NGC\,6819, M67 and NGC\,188 have been obtained with the FIES spectrograph at the Nordic Optical Telescope. The M67 target has been used to benchmark the analysis, as it is a well studied cluster with asteroseismic data from the K2 mission. For the other two clusters we have obtained higher quality data than what had been analyzed before, which allows us establish their chemical composition more securely. Using a line by line analysis, we tested several different combinations of line lists and programs to measure equivalent widths of stellar absorption lines to characterize systematic differences within the same spectroscopic method.}{The obtained parameters for the benchmark star in M67 vary up to $\sim$\SI{170}{K} in effective temperature, $\sim$\SI{0.4}{dex} in $\log g$ and $\sim$\SI{0.25}{dex} in [Fe/H] between the different tested setups. Using the combination of equivalent width measurement program and line list that best reproduce the inferred surface gravity from asteroseismology, we determined the atmospheric parameters for the three stars and securely established the chemical composition of NGC\,6819 to be close to solar, [Fe/H]=-0.02$\pm$\SI{0.01}{dex}.}{We have highlighted the significantly different results obtained with different combinations of line lists, programs and atomic parameters. The results emphasize the importance of benchmark stars studied with several different methods to anchor spectroscopic analyses.}


\maketitle

\section{Introduction}

Several large spectroscopic surveys such as RAVE \citep{Steinmetz2006}, APOGEE \citep{Majewski2015}, the Gaia-ESO survey \citep[GES,][]{Gilmore2012,Randich2013} and GALAH \citep{DeSilva2015} are being carried out to better understand the structure, kinematics and evolution of stars and our Galaxy. Massive amounts of data are obtained every day that require a swift, precise, and accurate analysis in order to support the scientific goals of the aforementioned surveys. To achieve this, pipelines have been constructed that analyze the data in a homogeneous way, e.g. ASPCAP \citep{Perez2015} for the case of APOGEE. These pipelines are normally tested against {\it benchmark stars}: stars that have been observed and analyzed by more than one method and have extremely high quality data. In the case of atmospheric parameters, examples of benchmark stars can be targets that have either abundances determined to a high level of precision \citep[e.g.,][]{Jofre2014}, effective temperatures measured by interferometry \citep[e.g.,][]{Casagrande2014}, and ideally also an independent estimate of surface gravity. In the latter case, the advent of asteroseismology as a tool for Galactic archaeology has opened a new window of opportunities via its synergies with large spectroscopic surveys by providing precise measurements of $\log\,g$ for thousands of stars across the Milky Way \citep[see e.g.,][just to name a few examples]{Miglio2013,Casagrande2016,SilvaAguirre2018,Pinsonneault2018}

Many different methods are available for spectroscopic analyses of optical stellar spectra, varying from fitting synthethic spectra to observed spectra to classical equivalent width methods. For each of these there are also different options for programs, where e.g. Spectroscopy Made Easy \citep[SME,][]{Valenti1996} is widely applied for synthetic fitting using a chi-square minimization algorithm and MOOG \citep{Sneden1973} is widely used for the classical equivalent width approach. It is well known that these different options do not always yield the same result and the extent of the differences can be significant. For a comparison of spectral synthesis methods, which we will not explore further here, see e.g. \citet{Lebzelter2012}. For a comparison of equivalent width studies in the Gaia-ESO framework, see e.g. \citet{Smiljanic2014}, where agreement between multiple determinations of atmospheric parameters for the same star are better than \SI{82}{K}, \SI{0.19}{dex} and \SI{0.10}{dex} for $\rm T_{eff}$, $\log g$ and $\rm [Fe/H]$, respectively. For larger comparisons of both spectral synthesis methods and equivalent width methods see e.g. \citet{Jofre2014,Hinkel2016} where more detailed discussions on this and efforts to understand the causes of variations between stellar abundance measurement techniques can be found. Both studies have a list of stellar spectra analyzed with different methods and with and without restrictions on stellar parameters and line list. \citet{Casamiquela2017} studied the differences between spectral synthesis and equivalent widths by analyzing data of red clump stars in several open clusters and found an offset between the two methods in metallicity of $0.07 \pm \SI{0.05}{dex}$. Moreover, the choice of absorption lines can highly affect the result. Even when choosing only lines that have the best measured atomic parameters and are not blended to any level of significance, strikingly different results in atmospheric parameters can be obtained, see e.g. \citet{Doyle2017} who find variations up to \SI{50}{K}. This paper also aims at highlighting this issue.

To perform an accurate spectroscopic analysis, it is necessary to know the level of systematic uncertainties incurred when a particular selection of e.g. synthetic fitting method or equivalent width measurement program is made. This can be achieved by a detailed comparison of the results provided by these combinations when analyzing the same spectra, if additionally an empirical measurement of a stellar property exists. For example \citet{Smiljanic2014} and \citet{Casamiquela2017} used cluster membership as an independent constraint to evaluate the results. In this paper, we take one step further in this direction by not only including cluster studies but also asteroseismology in comparison with spectroscopic results obtained when varying one source of systematic uncertainty at the time.

In this paper, we have chosen the method of measuring equivalent widths of individual absorption lines and requiring excitation and ionization equilibria to hold when performing a differential anlysis. We compare several combinations of line lists and programs to measure equivalent widths in order to understand the systematic uncertainties within the line by line spectroscopic analysis. 

We carry out a detailed fundamental parameter and abundance analysis of three targets, one in each of the three open clusters (NGC\,6819, M67 and NGC\,188), based on high-resolution and high signal-to-noise (S/N) spectroscopic data from FIES (FIbre-fed Echelle Spectrograph) at the Nordic Optical Telescope (NOT). The targets have been selected to be in the same evolutionary phase (helium-core-bruning) to make the analysis as homogeneous as possible. In the case of NGC\,6819 and M67 there is also asteroseismic data available from the \textit{Kepler} \citep{Borucki1997} and K2 missions \citep{Howell2014}. Asteroseismology provides $\log g$ values with much higher precision than what can normally be achieved with spectroscopy, which can greatly help constrain the spectroscopic analysis \citep[see e.g.][where asteroseismology has been used to calibrate the spectroscopic surface gravity by RAVE and APOGEE, respectively]{,Valentini2016,Pinsonneault2015}. It has however been shown that when doing a fully differential spectroscopic analysis of very high quality data, precision levels comparable to asteroseismology are achievable. \citet{Nissen2015} did a differential analysis of solar twins to the Sun with very high quality spectra, $R=115,000$ and S/N levels above 600, where an average uncertainty on the surface gravity of only \SI{0.012}{dex} is obtained. This work was also compared and found to be in excellent agreement with a similar analysis by \citet{Ramirez2014b} who achieved a surface gravity precision of \SI{0.019}{dex} for spectra of $R=83,000$ and S/N levels above 400. It is however important to note that this high level of precision is only achievable between stars of similar spectral type.

We have chosen stars in open clusters, because they can be assumed to originate from the same molecular cloud and are therefore expected to have similar chemical composition and age. This also implies that by studying one confirmed member of a cluster, we get information about an entire population of stars. Open clusters have been the subject of extensive studies throughout the years, and therefore our results can be compared to literature values. This also has implications for other fields in astrophysics; clusters can be used as calibrators when studying galaxies, distances in the Universe and age scales for stars. In order to exploit this fully, we need precisely determined properties of cluster stars involving different observing techniques to secure the accuracy. 

The data used in our analysis for both NGC\,6819 and NGC\,188 is of higher quality, resolution and signal-to-noise (S/N), than previously used. To our knowledge, there is no high-resolution data (R>65,000) that also have S/N>100 for NGC\,188 in the literature \citep[see e.g][for previous studies of this cluster]{Randich2003,Friel2010,Casamiquela2017}. Especially, the higher quality data will allow us to better determine the metallicity of NGC\,6819, on which there still is not a full consensus in the literature even though it is an otherwise very well studied cluster by asteroseismology because it is in the \textsl{Kepler} field. Previous determinations of the metallicity have been done by e.g. \citet{Bragaglia2001} and \citet{Lee-Brown2015} who, respectively, analyzed three targets with $R \sim 40,000$ yielding [Fe/H] = $0.09 \pm $\SI{0.03}{dex} and 200 targets with $R \sim 13,000$ yielding [Fe/H] = $-0.02 \pm $\SI{0.02}{dex}.

\section{Observations}
\label{sec:obs}

We have gathered data for confirmed members \citep{Hole2009,Yadav2008,Stetson2004} in the helium core-burning phase, one in each of the three open clusters; NGC\,6819, M67 and NGC\,188. M67 is a very well studied nearby solar-like cluster, which has also been observed with the K2 mission, and that target will serve as a benchmark in the analysis. 
All targets are seen to be in the helium-core-burning evolutionary phase from color-magnitude diagrams, see Fig.~\ref{fig:cmd}. Futhermore the two targets in M67 and NGC\,6819 are also confirmed by asteroseismology to be in the helium-core-burning phase \citep{Corsaro2012,Stello2016}. 

The observations and data reduction are presented and described in more detail in section 2 and Table A.1 of \citet{Slumstrup2017}. Briefly, the spectra were obtained during period 47 and 51 at the Nordic Optical Telescope (NOT) with the FIbre-fed Echelle Spectrograph (FIES). The covered wavelength range is \SI{3700}{\angstrom}-\SI{7300}{\angstrom}, with high resolution ($R=67,000$) and high S/N (above 100), which was estimated from the rms variation of the flux in a region around \SI{6150}{\angstrom} for each spectrum. The co-added spectrum for each star was normalized order by order using RAINBOW\footnote{\scriptsize \url{http://sites.google.com/site/vikingpowersoftware}}, which uses synthetic spectra with atmospheric parameters close to the expected for the target to guide the user to identify continuum points in the observed spectrum. These are then fitted with a spline function, which the spectrum is divided by to normalize it. After this careful normalization, the orders are merged and the spectrum is shifted to laboratory wavelength to secure the line identification in the further analysis.

\section{Prior information}
\label{sec:prior}

To have a starting point for the spectroscopic analysis, we use photometry and asteroseismology to give first guesses on effective temperature, $T_\text{eff}$, surface gravity, $\log g$, and metallicity, [Fe/H].

\begin{figure*}
\centering
\includegraphics[width=.30\textwidth]{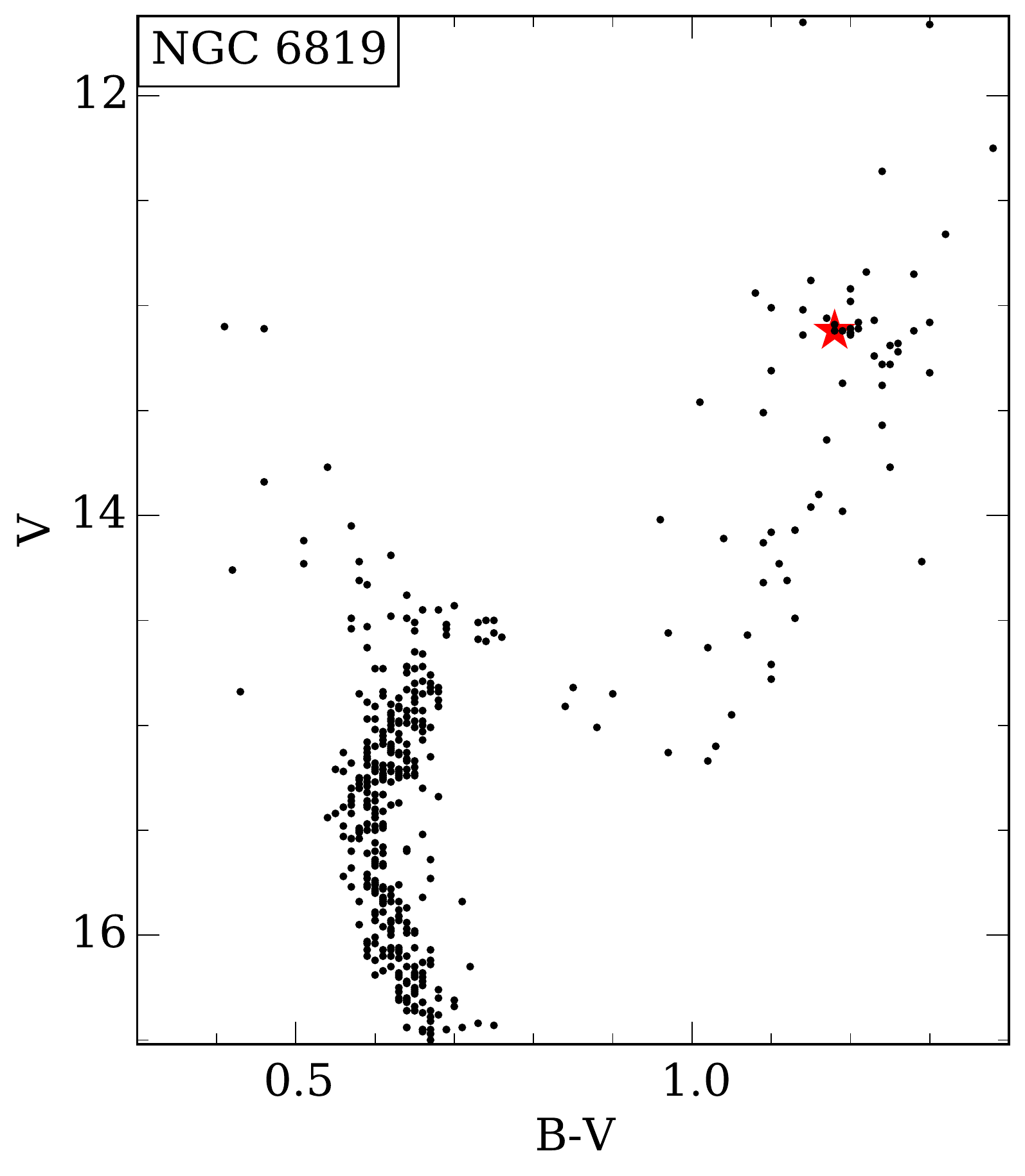}
\includegraphics[width=.30\textwidth]{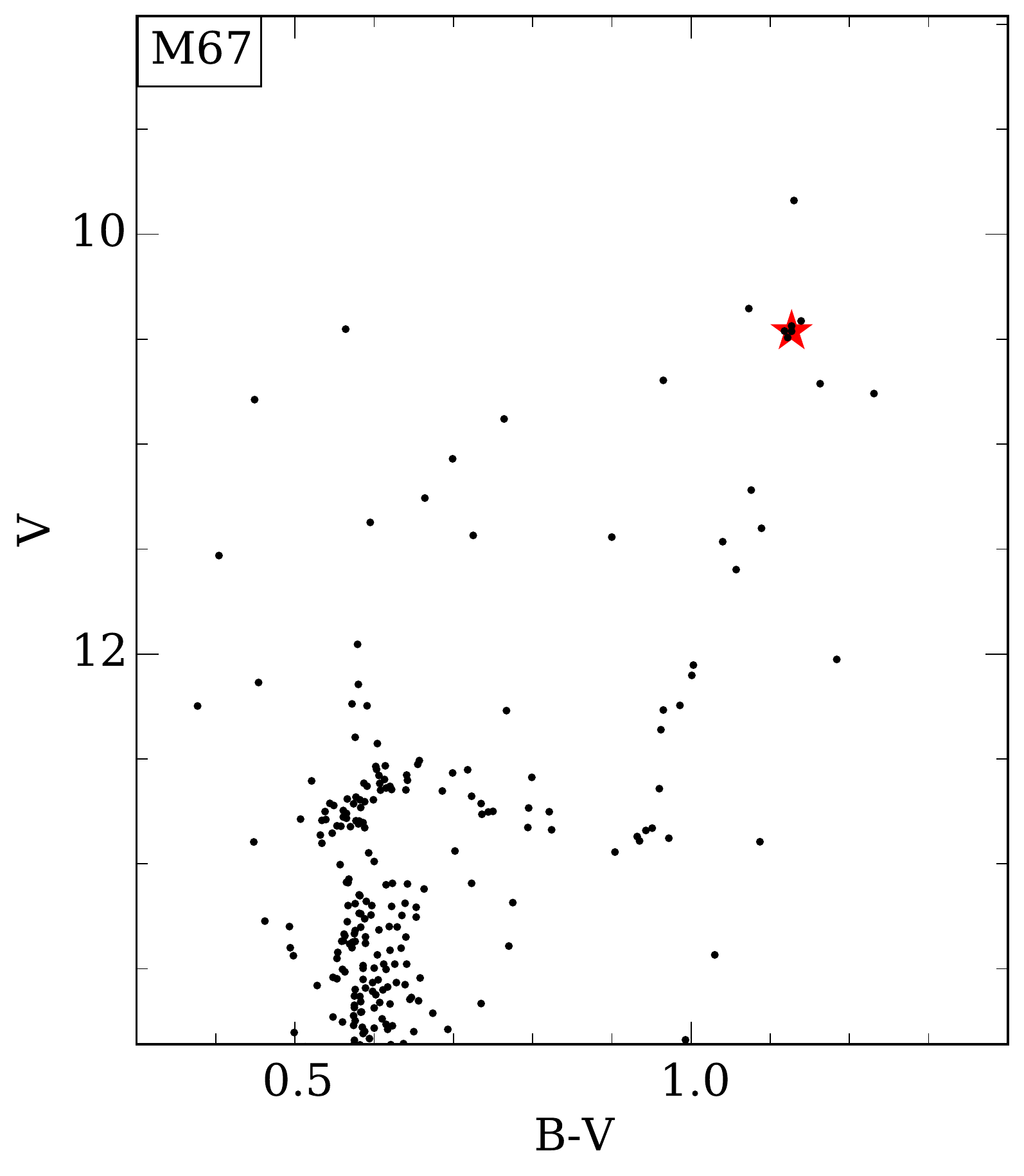}
\includegraphics[width=.30\textwidth]{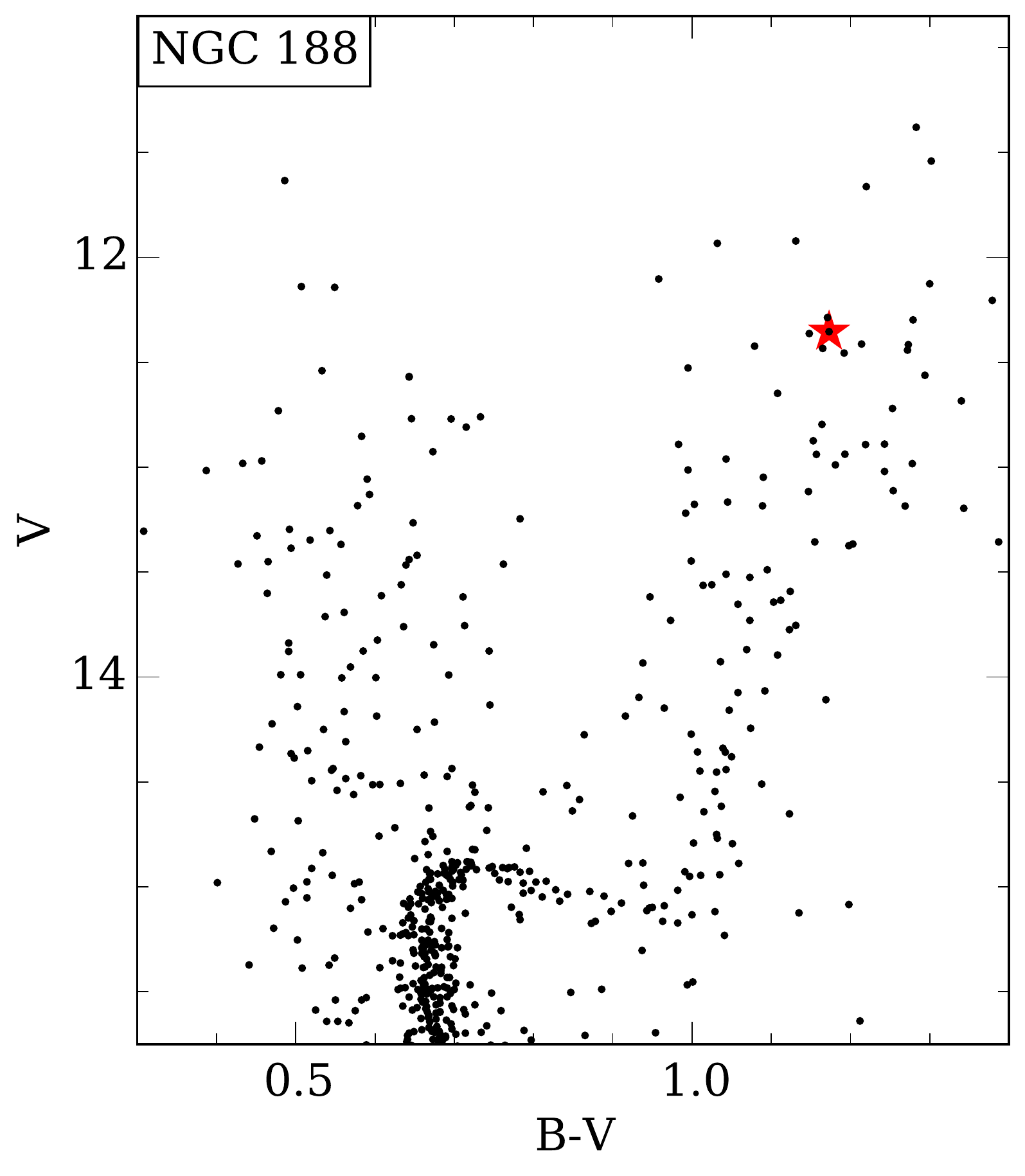}
\caption{From left to right: The CMDs for NGC\,6819 and M67 with photometry from \citet{Hole2009, Yadav2008} respectively and lastly a CMD for NGC\,188 with photometry from the \citet{Stetson2004} standard fields. The target star for each cluster is marked with a red star.}
\label{fig:cmd}
\end{figure*}

First guesses of effective temperature for the three targets were calculated with the metallicity dependent color-temperature relations by \citet{Ramirez2005}, which are based on the infrared flux method. The optical photometry used is presented in Fig.~\ref{fig:cmd} and the infrared photometry is from the 2MASS catalog \citep{Skrutskie2006}. The reddening values \citep[presented in Table 1 of][]{Slumstrup2017} used for these calculations are from \citet{Rosvick1998,Taylor2007,Meibom2009} for NGC\,6819, M67 and NGC\,188 respectively. Results for each filter combination are presented in Table~\ref{tab:coltemp}.
\begin{table}\small
 \centering
	\caption{\textsl{Temperatures from temperature-color calibrations.}}
	\begin{threeparttable}
		\begin{tabular}{l c c c c c c}
			\toprule
			Target & $B-V$ & $V-I$ & $V-K$ & $V-J$ & $V-H$ & avg\\
			\midrule
			NGC6819 & 4749 & -    & 4685 & 4659 & 4698 & 4698 $\pm$ 33\\
			M67     & 4643 & 4736 & 4697 & 4763 & 4749 & 4718 $\pm$ 43\\
			NGC188  & 4650 & 4610 & 4618 & 4651 & 4656 & 4637 $\pm$ 19\\
			\bottomrule
		\end{tabular}
	\end{threeparttable}
	\label{tab:coltemp}
 \end{table}
The initial guesses on the temperatures were taken as averages of the different filter combinations for each target (last column of Table~\ref{tab:coltemp}). The uncertainties on the photometric temperatures are only the scatter between the different color-temperature relations for each target and do not take into account the uncertainty in the relations themselves and should only be taken as a guide.

With the frequency of maximum power, $\nu_\text{max}$ from asteroseismology and $T_\text{eff}$ from photometry, the surface gravity can be calculated using the scaling relation (\cite{Brown1991}, \cite{Kjeldsen1995}):
\begin{align}
\log g = \log \left( \left( \frac{\nu_\text{max}}{3100} \right) \cdot \left( \frac{T_\text{eff}}{5777} \right) ^{1/2} \right) + 4.44 \ .
\label{eq:seis}
\end{align}
This relation is extrapolated from the Sun. Since we apply it here for giants, deviations should be expected, which have been quantified in the literature with, e.g. binary studies, parallaxes and comparisons to interferometry. Asteroseismic masses have been shown to be accurate to better than 8\% \citep{Stello2016,Miglio2016,Brogaard2017} and radius better than 4\% \citep{Huber2012,SilvaAguirre2012,White2013,Huber2017,SilvaAguirre2017,Sahlholdt2018}. Assuming a stellar mass of \SI{1.2}{M_\odot} and radius of \SI{10}{R_\odot}, this translates to an accuracy on $\log g$ better than \SI{0.04}{dex}. Even with only three months of data for the M67 target with K2, the precision is still better than what has been achieved for giants with normal spectroscopic methods in the literature that often suffer from systematic errors on the order of \SI{0.2}{dex} \citep[e.g.][to name a few]{Bruntt2011,Thygesen2012a,Heiter2015}
, the exception being a strictly differential analyses described in the introduction. This provides the strongest constraint to the spectroscopic analysis.  
NGC\,6819 is in the Kepler field with asteroseismic measurements available from extensive photometric time series \citep{Corsaro2012,Handberg2017}. M67 is in field 5 of the K2 mission, which also provides asteroseismology but with lower S/N \citep{Stello2016}, as only three months of time series data has been acquired compared to the four year extent of the nominal Kepler mission.

There are no oscillation measurements available for NGC\,188, but because there is a binary system near the main sequence turn-off in this cluster it is still possible to calculate an estimate of the surface gravity from photometry \citep{Nissen1997} to a higher precision than for field stars with no strong mass constraints:
\begin{align}
\log \frac{g}{g_\odot} = 4.44 + \log  \frac{M}{M_\odot} + 4 \cdot \log \frac{T_\text{eff}}{5777}  + 0.4 \cdot \left( V - 3.1E \left( B-V \right) \right) \nonumber \\ 
+ 0.4 \cdot BC + 2 \cdot \log \left( \frac{1}{d} \right) + 0.12  \ ,
\label{eq:logg}
\end{align}
where $M$ is the mass. $M_\text{NGC\,188} = 1.1 M_\odot$ is found by using a turn-off mass in the cluster of $1.1 M_\odot$ \citep{Meibom2009} and assuming the evolution and mass loss after the turn-off advances as in NGC\,6791 the mass difference between the turn-off and red clump will correspond to the mass loss on the red giant branch \citep{Brogaard2012,Miglio2012}. $V$ is the $V$-band magnitude, $BC$ is the bolometric correction and $d$ is the distance calculated as
\begin{align}
d = 10^{\left( 1 + \frac{\mu - 3.1 \cdot E \left( B-V \right)}{5} \right)} \ .
\end{align}
The apparent distance modulus in the $V$-band, $\mu = (m-M)_V$, is the difference between the apparent and absolute magnitude. The reddening and distance modulus for NGC\,188 are from \citet{Meibom2009} and presented in Table 1 of \citet{Slumstrup2017}.
An empirical relation for the bolometric correction in Eq. \ref{eq:logg} applicable for stars with $T_\text{eff}$~<~\SI{5012}{K} is given in \cite{Torres2010}:
\begin{align}
BC = a + b \left( \log T_\text{eff} \right) + c \left( \log T_\text{eff} \right)^2 + d \left( \log T_\text{eff} \right) ^3 \ ,
\end{align}
where $a,b,c,d$ are coefficients given in the article. The surface gravities from asteroseismology and photometry for the three targets are: NGC\,6819: $\log g = 2.55 \pm$\SI{0.02}{dex}, M67: $\log g = 2.48 \pm$\SI{0.06}{dex}, NGC\,188: $\log g = 2.45 \pm$\SI{0.14}{dex}.

\section{Comparison of methods}

To achieve a robust result, we did several realizations of the spectroscopic analysis using different combinations of line lists and equivalent width measurement programs and compared the outcome to determinations from asteroseismology and photometry. Due to the high accuracy of the asteroseismic $\log g$ as mentioned in Sec.~\ref{sec:prior}, we have used this as the main calibrator throughout the analysis. In the following section we describe the analysis carried out in detail, where we have used the M67 target as a benchmark and SPECTRUM \citep{Gray1994} to calculate atmospheric parameters from equivalent widths (see discussion in Sec.~\ref{sec:param}).

When asteroseismology is available the $\log g$ value can be held fixed in the spectroscopic analysis, which is often the chosen approach when possible. This can result in ionization equilibrium between FeI and FeII not being reached in the spectroscopic analysis, but In most cases it will not have a large impact on the effective temperature because excitation balance, which is used to obtain $T_\text{eff}$, depends on the excitation potentials and abundances of FeI lines, which are only mildly sensitive to pressure changes in the atmosphere.

Instead of fixing the $\log g$ value during our spectroscopic analysis, we used it as a reference when choosing our preferred combination of line list and equivalent width measurement program to make the spectroscopic results self consistent and in agreement with asteroseismology. \citet{Doyle2017} investigated the effect of fixing $\log g$ in the spectroscopic analysis of a set of FGK stars, that already had accurate parameters determined from other methods, and found that fixing the surface gravity did not improve the precision on the other spectroscopic parameters, they find an average difference in determined $T_\text{eff}=3\pm$\SI{13}{K}. This is however in contrast to \citet{Hawkins2016} and \citet{Valentini2016} who both find better precision on the other atmospheric parameters when fixing the $\log g$ to the asteroseismic.

\subsection{Line lists}
\label{sec:line}

To test the impact of choosing different combinations of absorption lines, we adopt one equivalent width measurement program (DAOSPEC, see discussion in Sec.~\ref{sec:program}) .
When choosing a line list, several considerations have to be made. The lines should not be blended, they should have well determined atomic parameters, they should cover a range of excitation potentials to get a robust effective temperature determination. Furthermore, they should preferably be weak enough to be on the linear part of their curve of growth, yet strong enough to yield a significant determination, again covering a range of line strengths. 

We ensured our lines were not blended by evaluating each individual transition for our range of atmospheric parameters with the online tool for the Vienna Atomic Line Database \citep[VALD,][]{Piskunov1995}. Based on our S/N values of 100-150, we decided that a possible blend should be stronger than 5\% to be considered significant. This however only concerns atmospheric blends. A separate issue is that of telluric blends, which are in the rest frame of the Earth and can therefore not be evaluated with a tool like VALD. In order to avoid telluric blends, we chose to avoid parts of the spectrum with well-known telluric absorption lines.

\begin{figure*}
\centering
\includegraphics[width=.8\textwidth]{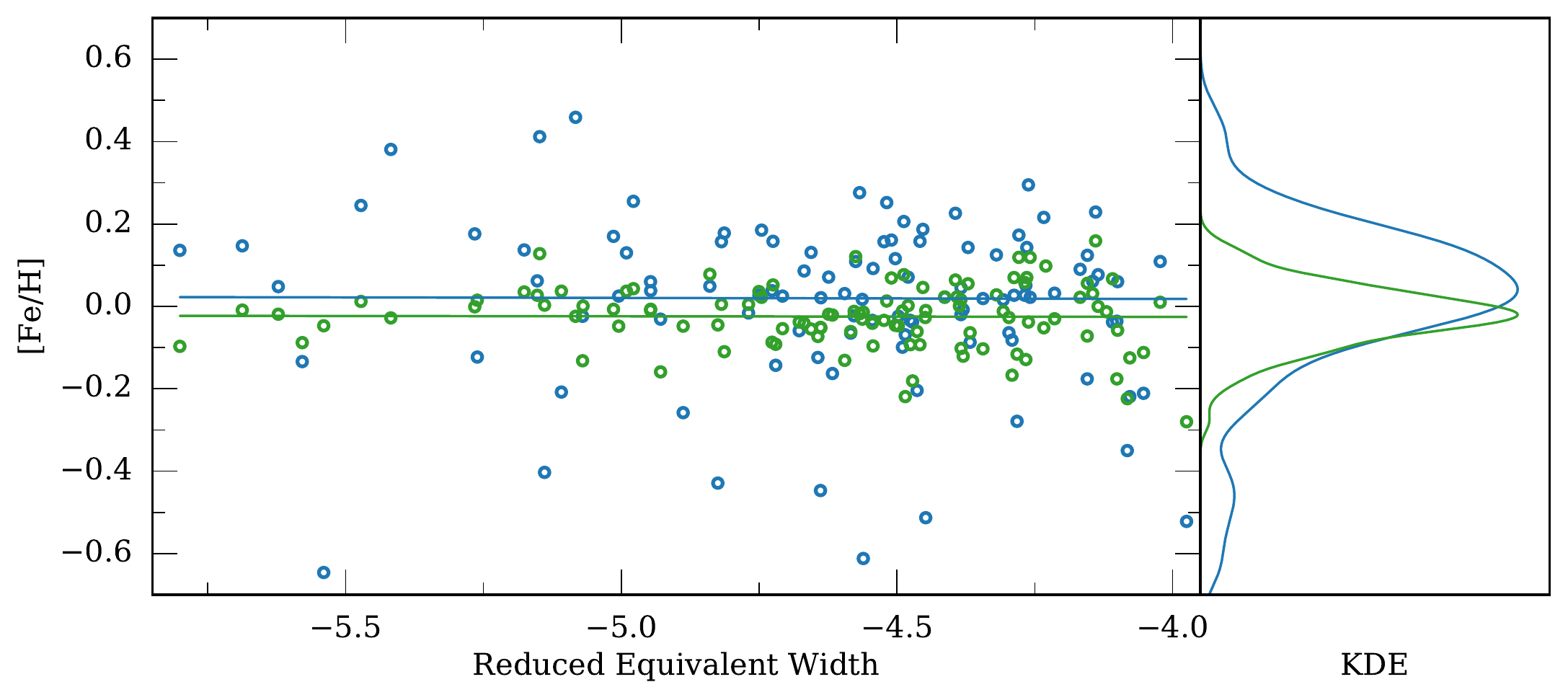}
\caption{Comparison of results using DAOSPEC and the line list based on \citet{Carraro2014a} for laboratory (blue) vs astrophysical (green) $\log \-gf$ values for the M67 target. The right plot is a KDE illustrating the scatter in the two samples.}
\label{fig:labvsastro}
\end{figure*}

We were careful in choosing absorption lines with only reliably determined atomic parameters, but there are however still lines that do not yield the expected element abundances, e.g. Fe lines in the Sun not giving solar metallicity when using solar atmospheric parameters. This is the reason to use the so-called astrophysical oscillator strengths, $\log \-gf$ values as an attempt to minimize effects from our lack of knowledge on atomic parameters. These are calibrated to the Sun by analyzing a solar spectrum taken with the same instrumental setup as the data and then adjusting the $\log \-gf$ values until each absorption line yields the expected solar chemical abundance. This can result in an increase on the precision of the metallicity by a factor of two when using calibrated values compared to laboratory values. It is however important to note that these values will give more precise results by construction, but in doing so a bias can possibly be introduced in the analysis and care must be taken.
Figure~\ref{fig:labvsastro} shows this comparison for two sets of results for the M67 target using the same line list and program for measuring equivalent widths, where one considers laboratory $\log \-gf$ values and the other astrophysical values. The kernel density estimator (KDE) plot on the right shows that not only is the scatter larger for the laboratory values (\SI{39}{K} compared to \SI{18}{K}), but also the resulting atmospheric parameters change: the chemical composition, effective temperature and surface gravity increase when using laboratory values by $\Delta$[Fe/H] = +0.05, $\Delta T_\text{eff}$ = \SI{+100}{K} and $\Delta \log g = +0.27$, respectively (the difference between the two bottom results in Fig.~\ref{fig:comp_res}). The value for the surface gravity of the M67 target is in poor agreement with the results from asteroseismology ($\log g_\text{seis} = 2.48 \pm 0.06$ as shown in Fig.~\ref{fig:logg}), and photometry ($\rm T_{eff,phot} = 4718 \pm \SI{43}{K}$ as shown in Table~\ref{tab:coltemp}). For the other two targets, NGC\,6819 and NGC\,188, the trend goes in the same direction but is more pronounced: $\Delta$[Fe/H] = +0.02, $\Delta T_\text{eff}$ = \SI{+160}{K} and $\Delta \log g = +0.35$ for NGC\,188 and $\Delta$[Fe/H] = +0.03, $\Delta T_\text{eff}$ = \SI{+170}{K} and $\Delta \log g = +0.38$ for NGC\,6819. We therefore chose to use the calibrated $\log \-gf$ values in our analysis. \citet{Doyle2017} finds similar trends when comparing laboratory and solar calibrated $\log gf$ values, that [Fe/H] is systematically higher with laboratory values, however their discrepancy on effective temperature is much smaller, <\SI{10}{K}.

\begin{figure*}
\centering
\includegraphics[width=.45\textwidth]{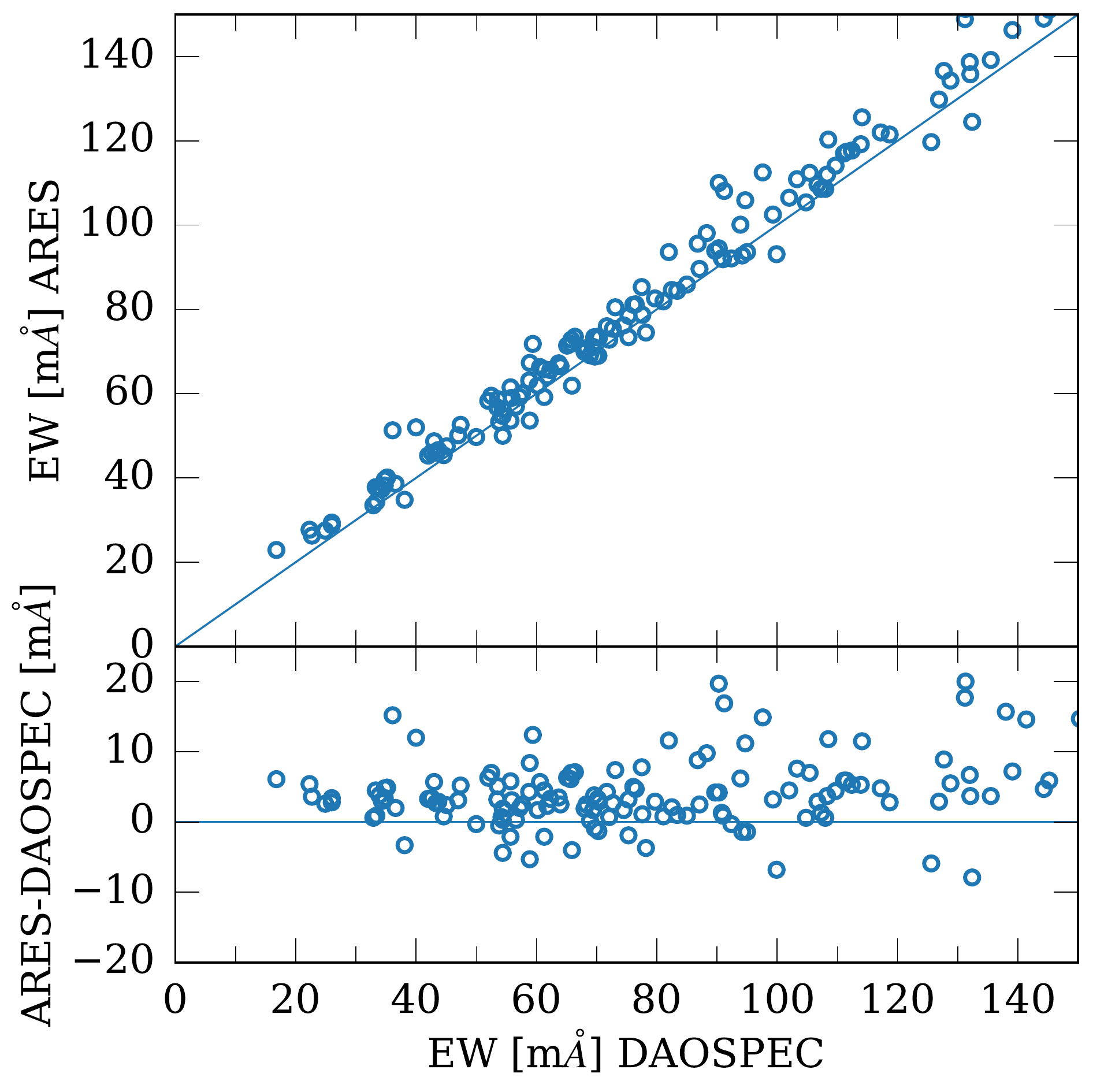}
\includegraphics[width=.45\textwidth]{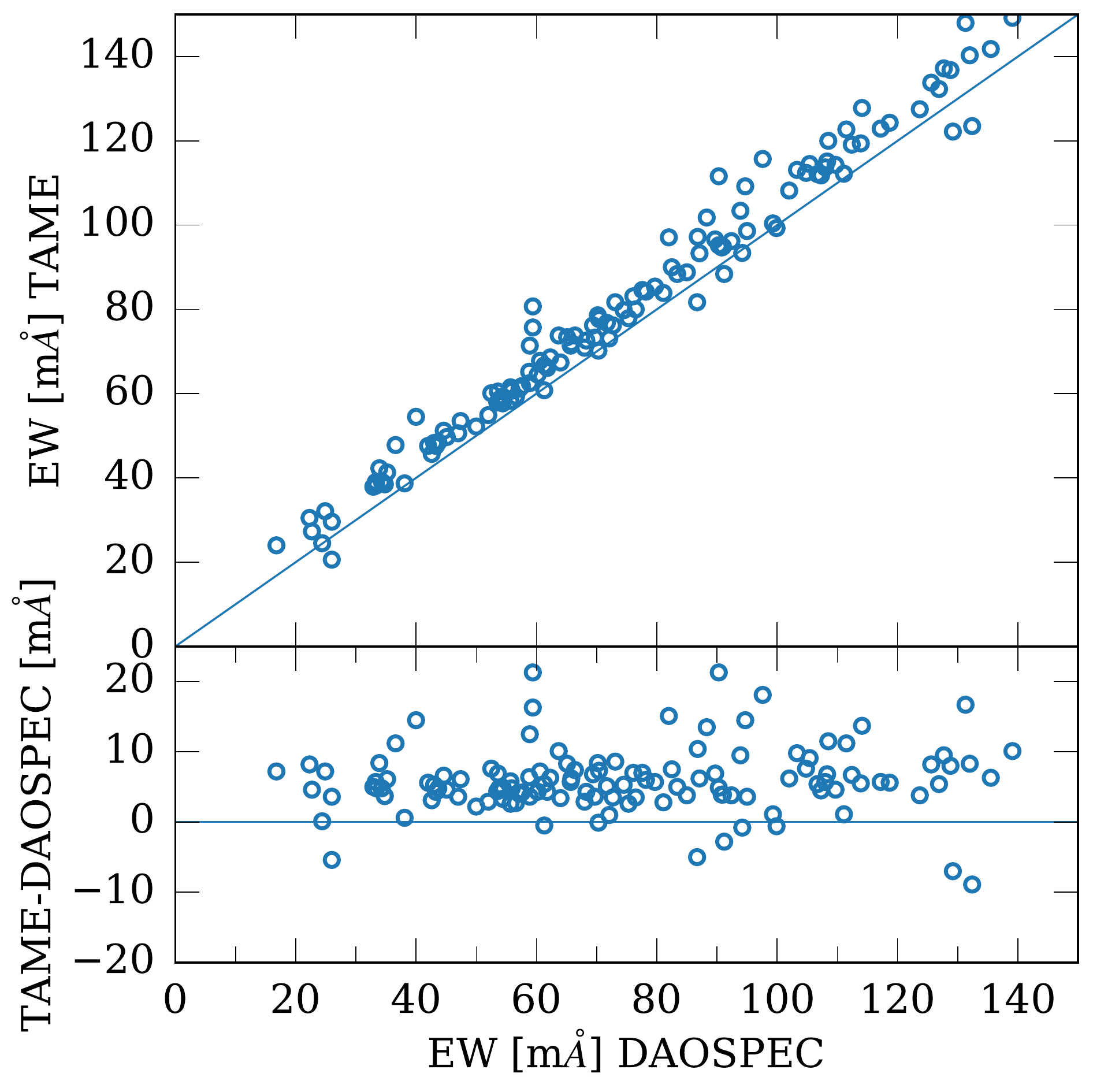}
\caption{Comparison of EWs for the M67 target measured with three routines. Left: ARES vs DAOSPEC with an average offset of \SI{4.8}{\milli\angstrom} Right: TAME vs. DAOSPEC with an average offset of \SI{5.9}{\milli\angstrom}. In the spectroscopic analysis, Fe lines above \SI{100}{\milli\angstrom} are discarded.} 
\label{fig:comp_ew}
\end{figure*}

As mentioned earlier in this section, the absorption lines have to preferably be weak enough to be on the linear part of the Curve of Growth (COG), which describes how the equivalent width of a line changes as the number of absorbers increase due to a change in a specific atmospheric parameter, e.g. abundance of the element in question, $T_\text{eff}$ or $\log g$. For the three stars analyzed here, the linear part of the COG corresponds to lines with an equivalent width up to about \SI{100}{\milli\angstrom}. Figure~\ref{fig:comp_ew} shows a comparison between equivalent width measurements of the same lines but with different programs (more on this in Sec.~\ref{sec:program}). Especially in the left plot of Fig.~\ref{fig:comp_ew} it becomes apparent why stronger lines should be left out when possible: the scatter increases significantly for the stronger lines (around $\sim$\SI{90}{\milli\angstrom} and above). For instance, when including the stronger lines for the M67 target with astrophysical $\log gf$ values the atmospheric parameters changed as well: $T_\text{eff}$ goes from \SI{4680}{K} to \SI{4650}{K} ($\Delta T_\text{eff}$=\SI{-30}{K}), $\log g$ from \SI{2.43}{dex} to \SI{2.30}{dex} ($\Delta \log g$= \SI{-0.13}{dex}) and [Fe/H] from -0.03 to 0.00 ($\Delta$[Fe/H]=0.03). For some elements with only a few good absorption lines in the optical part of the spectrum for a given atmospheric parameter range, the requirement of \SI{100}{\milli\angstrom} can be difficult to reach. In our case, a few lines with higher equivalent widths than the \SI{100}{\milli\angstrom} limit were used for elements other than Fe due to the lower number of lines available.

The higher the S/N of the spectrum, the weaker the chosen lines can be because the depth of the line has to be significant compared to the noise level of the spectrum. Given our S/N levels, we set the lower limit at \SI{10}{\milli\angstrom}.

\subsubsection*{Golden lines}

\cite{Jofre2014} collected results for a set of Benchmark stars studied in the framework of the Gaia-ESO collaboration using many different pipelines to study systematic differences in the analysis. They created a list of "golden lines" (presented in Fig.~\ref{fig:linelists}), which are FeI and FeII lines all with well determined atomic data that are on the linear part of the COG for all of their benchmark stars. These include FGK dwarfs and FGKM giants. Measurements of the strength of all the lines in the golden line list by several different groups agree within 2$\sigma$. Using only the golden lines for the FGK giants, the final scatter on the metallicity results from the different pipelines is around \SI{0.08}{dex}. This is a good place to start when choosing lines and we tested this line list on the M67 target. It resulted in the atmospheric parameters $T_\text{eff}$=\SI{4820}{\kelvin}, $\log g$=\SI{2.85}{dex} and [Fe/H]=\SI{-0.02}{dex} (see Fig.~\ref{fig:comp_res}, which shows the results from all the different combinations of line lists and programs considered in our study). The temperature is significantly higher than predicted from photometry (see Table~\ref{tab:coltemp}), but the deciding part is the surface gravity that is much higher than the \SI{2.48}{dex} predicted from asteroseismology. Even though an uncertainty on $\log g$ from spectroscopy of 0.2-\SI{0.3}{dex} is common, a deviation from the asteroseismic $\log g$ of almost \SI{0.4}{dex} allows us to discard this line list for the targets analyzed here.

\subsubsection*{Other line lists}

\begin{figure*}
\centering
\includegraphics[width=\textwidth]{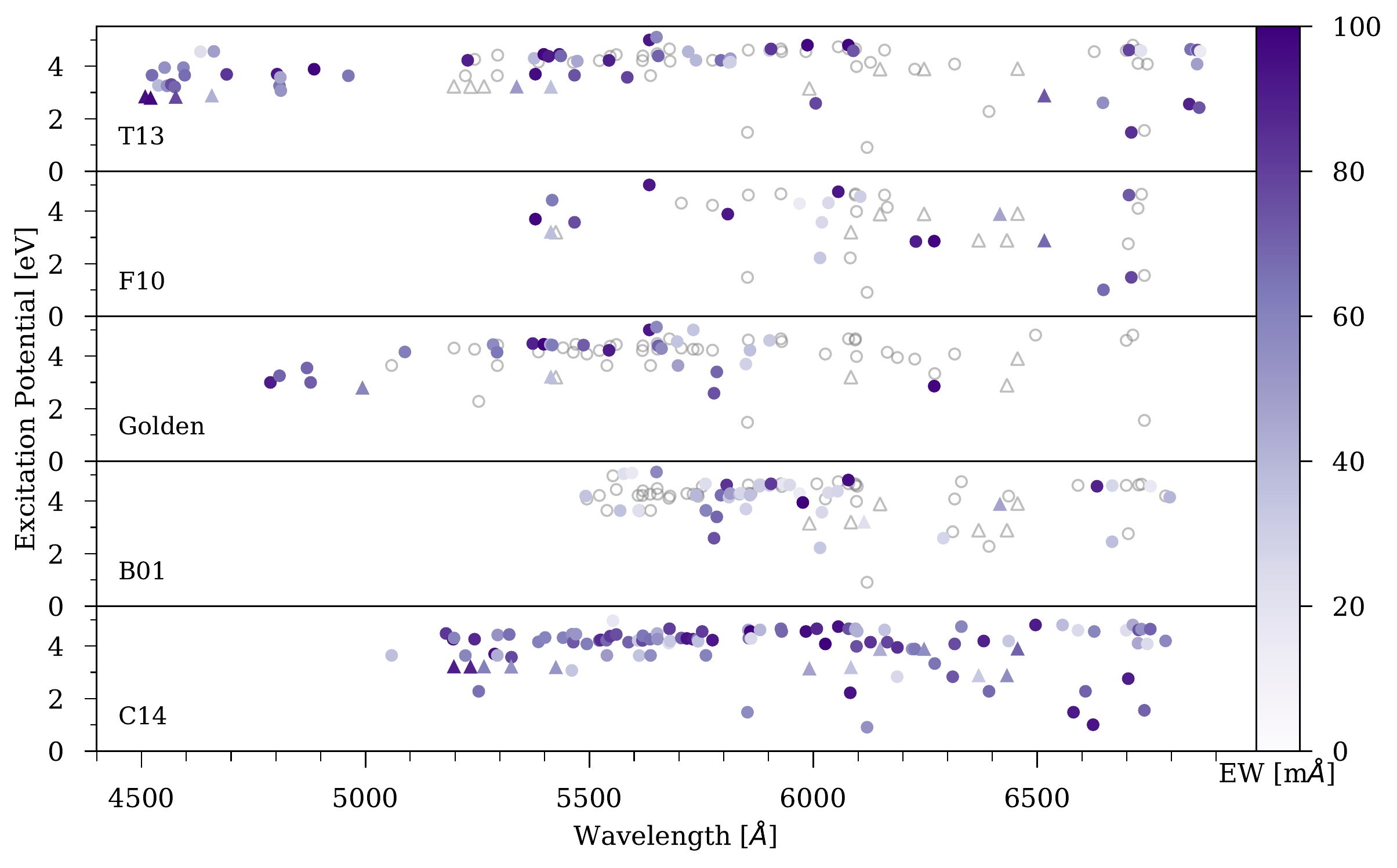}
\caption{The Excitation Potential of each line list plotted as a function of the wavelength. The color coding shows the equivalent widths measured with DAOSPEC. The circles are FeI lines and the triangles are FeII lines. The grey open circles and triangles in the top four panels are common lines with the bottom panel.}
\label{fig:linelists}
\end{figure*}
We tested other line lists as well based on \citet{Carraro2014a}, \citet{Bragaglia2001}, \citet{Friel2010} and \citet{Tsantaki2013} (from here on: C14, B01, F10 and T13 respectively). They are all presented in Fig.~\ref{fig:linelists} that shows the excitation potential of all the lines plotted as a function of wavelength color coded by the equivalent width. C14, B01 and F10 are all selected for cool giants, whereas T13 is for cool dwarfs. The lines in common from T13, F10, Golden and B01 with C14 are plotted with grey open symbols to highlight the differences between the line lists. From this it can be seen, that the main difference between them is the wavelength coverage. They all have mostly higher excitation potential lines (3 > eV > 5) and all span the entire range in line strengths. A few differences in the line lists stand out that could potentially trace the differences between the results: The F10 line list has the most even distribution of excitation potentials, which increases the stability of the results for $\rm T_{eff}$. It does however have significantly fewer lines than the other line lists, which overall decreases the precision on the results. The B01 and Golden line lists only have one and two lines respectively with excitation potential below \SI{2}{eV}, which makes the $\rm T_{eff}$ determination less accurate and because this has a strong effect on determination of the other parameters, the accuracy on these will also decrease. Care should however be taken with strong lines at low excitation potentials as these are more sensitive to 3D effects \citep[see][]{Bergemann2012}. The amount of FeII lines also has a strong effect on precision and accuracy of the surface gravity determinations, where especially the Golden line list stands out, as it has only six FeII lines below \SI{100}{\milli \angstrom}, which could be the explanation for the discrepancy with the asteroseismic surface gravity with this line list. A recent review on the accuracy and precision of stellar parameters is presented in \cite{Jofre2018}, where also the effects of line selection is discussed in more detail. Other differences on the obtained atmospheric parameters as a result of line selection could also be issues with the atomic parameters. Even though we have ensured all the lines have the best known atomic parameters, there could still be problems, as the theoretical calculations and laboratory measurements of transitions in iron is very difficult to carry out due to the complexity of the iron atom. Calibrating each line to the Sun would in principle remove these effects but due to the differences in atmospheric conditions between the Sun and our stars, this could add new problems. This is why it could also be a solution to calibrate to a well-known giant as e.g. Arcturus, but that would also add uncertainties to the analysis due to the uncertainty on the metallicity of Arcturus.

\subsubsection*{Chosen line list}

The results using the different line lists with DAOSPEC are presented in Fig.~\ref{fig:comp_res}. Based on the surface gravity determined from seismology, the final choice of Fe lines is comprised of the line list from C14 with astrophysical $\log gf$ values based on \citet{Grevesse1998} solar abundances and discarding lines stronger than $\sim$\SI{100}{\milli\angstrom}. 
The choice of non-Fe lines for the final determination of abundances is based on C14 and \citet{Nissen2015} plus a few additional lines as described in Sec.~3 of \citet{Slumstrup2017}. The C14 Fe line list with laboratory oscillator strengths yields a $\log g$ value within $\sim 3\sigma$ of the asteroseismic one, but with a much higher scatter as shown in Fig.~\ref{fig:labvsastro} and it was therefore discarded. The F10 line list also returns a $\log g$ value close to the asteroseismic one. Nevertheless, this line list has fewer usable absorption lines than the others (42 FeI and FeII lines compared to 109 with C14), which gives significantly larger uncertainties (see Fig.~\ref{fig:comp_res}) and a lower stability of the analysis (see discussion on this in Section~\ref{sec:uncertainty}).

\subsection{Equivalent Widths}
\label{sec:program}

The optimal way to measure equivalent widths is to do a manual line by line analysis, e.g. using routines such as \textit{splot} in IRAF. This can however be very time consuming since preferably one wants many lines for each target and therefore more automatic routines are often used. We tested three different programs that measure equivalent widths in a more automatic way with less user interaction;  DAOSPEC \citep{Stetson2008}, ARES \citep{Sousa2007} and TAME \citep{Kang2012}. Comparisons of equivalent width measurements between the three different programs for the M67 target is shown in Fig.~\ref{fig:comp_ew}.  

\begin{figure*}
\centering
\includegraphics[width=.95\textwidth]{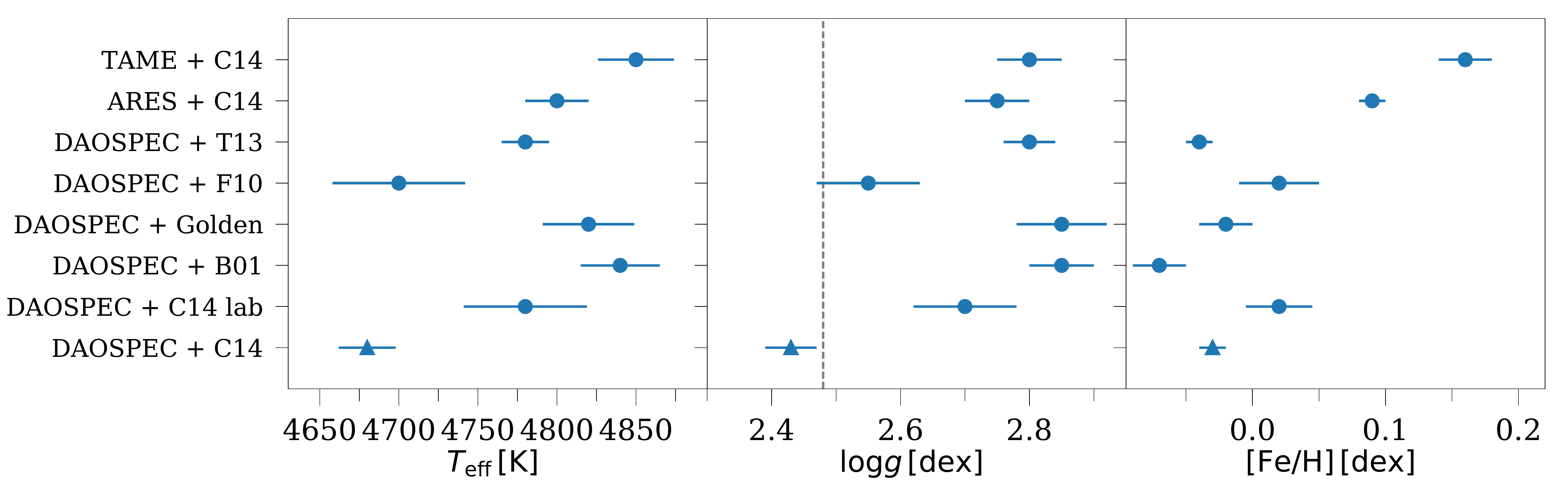}
\caption{Comparison of atmospheric parameters for the M67 target with different combinations of line list and program to measure equivalent widths. Left: $T_\text{eff}$ variations. Center: $\log g$ variations. Right: [Fe/H] variations. The final choice of line list and program is the bottom row marked with a triangle. The dashed line in the middle plot is the asteroseismic $\log g$, see Fig.~\ref{fig:logg} for a comparison.} 
\label{fig:comp_res}
\end{figure*}

TAME can be run in both with and without user interaction for the line fitting. The results we are comparing with here is with a manual user set continuum. When line profile fitting it is possible to choose between Gauss and Voigt profiles, where Voigt profiles should be used for the stronger lines. As mentioned previously, we have chosen not to work with stronger lines and therefor the comparison here is only for lines that can be fitted with a Gauss profile. TAME measures equivalent widths 5-\SI{7}{\milli\angstrom} higher than DAOSPEC. When using M67 as the test target the measured equivalent widths results in an effective temperature higher than that expected from photometry by \SI{130}{K} and a surface gravity deviating from that expected from asteroseismology by $\Delta \log g=+$\SI{0.30}{dex} (see Fig.~\ref{fig:comp_res}). This led us to discard this program. 

ARES fits lines with a Gaussian profile (for stronger lines, there are other options, but this was not relevant here) and does an automatic continuum placement locally around the absorption line aided by a parameter that depends on the S/N of the spectrum, which must be set by the user. In \citet{Sousa2008} a table is given with values of this parameter for different S/N levels. We varied this parameter to test its effect on the results in the range corresponding to 90<S/N<200. The impact on the effective temperature was on the order of $\sim\SI{30}{K}$ and  on the surface gravity it was $\sim$\SI{0.1}{dex}. These are not large variations and we therefore chose to only use the surface gravity as a calibrator due to the much stronger external constraint on this parameter. The result closest to the seismic $\log g$ value was obtained when using the parameter corresponding to a S/N between 100 and 125 (the estimated S/N for the M67 spectrum is around 135); the resulting value was however larger than the asteroseismic one by $\Delta \log g =+$\SI{0.20}{dex}. 

As mentioned in the previous section and shown on Fig.~\ref{fig:comp_ew}, the scatter on the line strength measurement increases for stronger lines (above $\sim$\SI{90}{\milli\angstrom}). For the targets in this project this was not very significant as we had enough well measured weaker lines and therefore did not have to include stronger lines, where perhaps fitting Voigt profiles instead of Gaussians become relevant. In other cases however it can be important to the point that removing the trend between the element abundance vs. reduced equivalent width to determine the microturbulence is impossible. The microturbulence will instead have to be calculated from a scaling relation, see e.g. \citet{Bruntt2012}. 
\citet{Sousa2007} did a comparison between DAOSPEC and ARES and found that on average, DAOSPEC measures the equivalent widths \SI{1.9}{\milli\angstrom} lower than ARES. In our study, the equivalent widths from ARES are 4-\SI{6}{\milli\angstrom} higher (left panel of Fig.~\ref{fig:comp_ew}), which is most likely mainly due to continuum placement, as ARES on average places the continuum ${\sim}$1\% higher than DAOSPEC for our spectrum. Exactly why the difference between ARES and DAOSPEC in our analysis is larger than that of \citet{Sousa2007} could be caused by a variety of reasons, e.g different line selection, different spectral types analyzed or the continuum placement, because as mentioned previously, the parameter used by ARES in the continuum placement for our spectrum, had to be set for a slightly lower S/N than the actual S/N of the spectrum.

\begin{figure}
\centering
\includegraphics[width=.9\columnwidth]{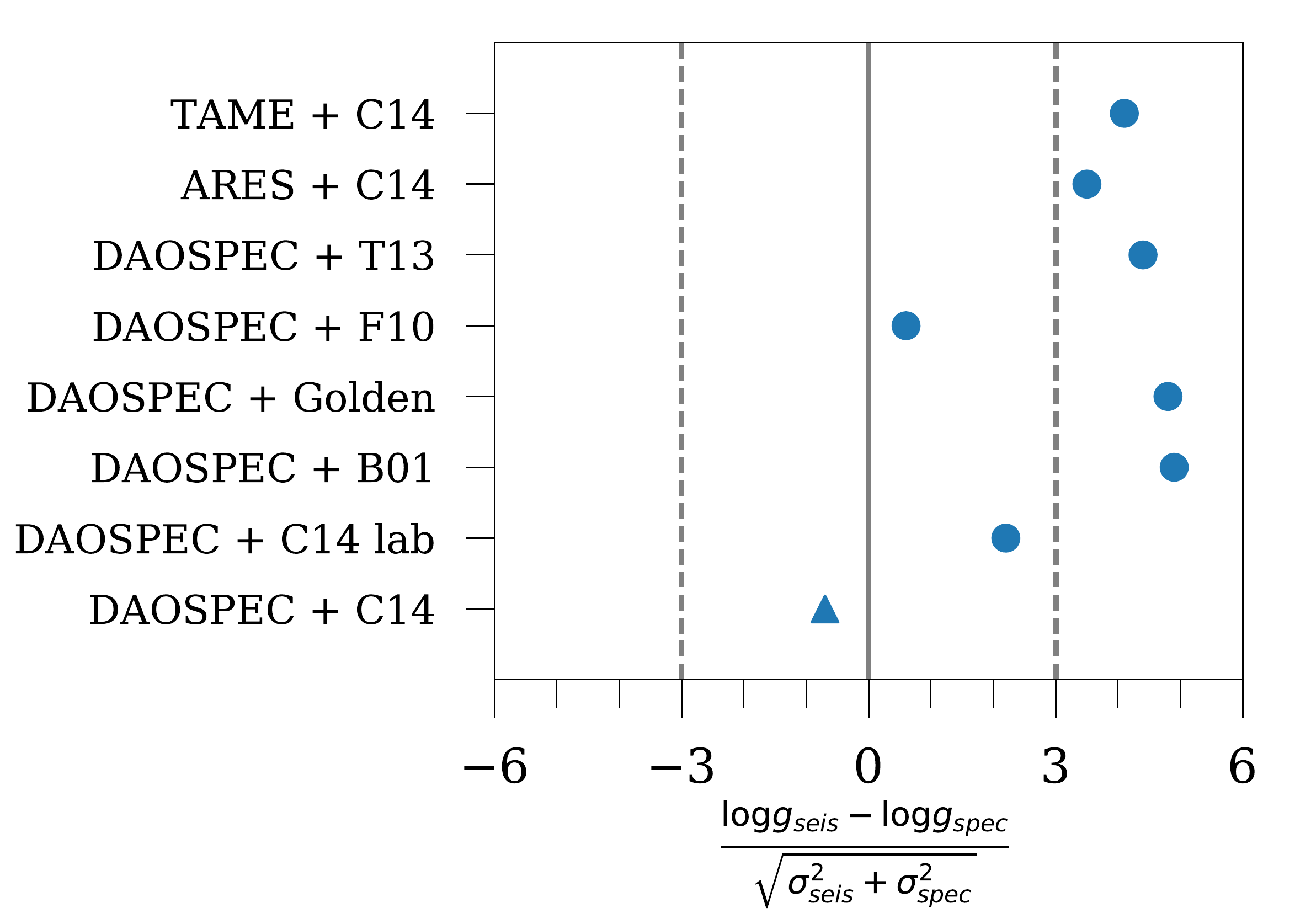} 
\caption{Deviation from the asteroseismic $\log g$ value for the M67 target for each method combination (center plot in Fig~\ref{fig:comp_res}). Full agreement with the asteroseismic $\log g$ is marked with the vertical line and the two dashed lines are $3\sigma$ deviations. The final choice of line list and program is the bottom row marked with a triangle.}
\label{fig:logg}
\end{figure}

The final choice of program fell on DAOSPEC, which also fits absorption lines with a Gaussian profile. The continuum is placed automatically around the absorption lines. It was chosen because as shown in Fig.~\ref{fig:comp_res} and \ref{fig:logg}, the measurement of absorption lines yielded results consistent with the more precise and accurate $\log g$ value from asteroseismology. A similar comparison can be performed with the effective temperature from photometry, which shows the same trend as Fig.~\ref{fig:logg}, but with systematically smaller deviations due to the larger uncertainty on the effective temperature from photometry. Even though the photometric effective temperature is less precise than the spectroscopic, it is reassuring that the results showing the best agreement with the asteroseismic $\log g$ are also the ones in closest agreement with the photometric $T_\text{eff}$. This further supports the choice of program and line list. 

An additional check of our measured equivalent widths was to apply our method to a solar spectrum. We used a reflected light spectrum of Vesta observed with the same setup as our targets (FIES in the high-resolution mode). 
We compare our results using DAOSPEC + C14 to solar equivalent widths measurements done by \citet{Scott2014}, and found very good agreement on the absorption lines we have in common. The average difference is \SI{-0.4}{\milli\angstrom} with a scatter of $\sim$\SI{3}{\milli\angstrom}. It should be noted, that since we have not measured the equivalent widths on the same exact solar spectrum, small differences can be expected. We found a few of our lines to be blended to an amount that makes them questionable to use in the solar case, which led us to also discard these lines in our analysis of the giants (N. Grevesse, private communication). After removing these lines, we end up with the following atmospheric parameters for the Sun: $T_\text{eff} =$ \SI{5770}{K}, $\log g = $ \SI{4.35}{dex}, $v_\text{mic} = $ \SI{0.90}{km/s}, [Fe/H] = \SI{-0.05}{dex}. The small difference to the actual solar atmospheric parameters could arise from the selection of absorption lines, which is tailored to more evolved stars than the Sun.


\section{Atmospheric parameters and abundances}
\label{sec:param}

From the comparisons described in the previous section, we chose an iron line list based on C14 and DAOSPEC to measure the equivalent widths. To calculate atmospheric parameters from the measured equivalent widths we use SPECTRUM \citep{Gray1994}. This routine carries out computations of atmospheric parameters under the assumptions of LTE and a plane-parallel atmosphere. SPECTRUM comes with auxiliary programs including ABUNDANCE, which is a routine for computing elemental abundances from equivalent widths of individual absorption lines using the COG method (as explained in Sec.~\ref{sec:line}).

One of the required inputs for a spectroscopic analysis is a stellar atmosphere model, where we have used a large grid of Kurucz-Castelli ATLAS9 models (\cite{Castelli2004}) interpolated to get the required atmospheric properties. These are 1D plane-parallel model atmospheres that divide the atmosphere of the star into many subsequent layers. Within each layer, the atmosphere is in hydrostatic equilibrium and energy transport is radiative. According to \cite{Asplund2005,Mashonkina2011} non-LTE effects are not relevant for FeII lines and mostly relevant for FeI lines in low metallicity stars ([Fe/H]<\-1.0). Thanks to the online tool INSPECT available at \url{www.inspect-stars.net} it is possible to check non-LTE corrections for individual lines given a line strength and atmospheric parameters. The data are from studies by \citet{Lind2012,Bergemann2012} and for the eight of our FeII and seven of our FeI lines where data are available, we get average changes in FeII of <\SI{-0.01}{dex} and in FeI of ${\sim}$\SI{0.01}{dex}. This is within the internal precision of our analysis and therefore we have chosen not to include non-LTE effects. 3D LTE effects on spectral line formation was discussed by \citet{Collet2007}, where fictitious lines at \SI{3500}{\angstrom} and \SI{5000}{\angstrom} were studied for different stellar atmospheres. For the atmosphere similar to the stars in this study ($\rm T_{eff}=\SI{4697}{K}, \log g=\SI{2.2}{dex}, v_{mic}=\SI{1.5}{km/s} and [Fe/H]=0.0$) the 3D-1D LTE corrections to the Fe abundance for the \SI{5000}{\angstrom} fictitious line vary between \SI{-0.1}{dex} and \SI{+0.2}{dex} depending on the excitation potential and the equivalent width of the line \citep[see Fig.~5 in][]{Collet2007}. We cannot apply these corrections here, as it could be anywhere in the range, also consistent with zero, but should however be kept in mind. Lines at \SI{6000}{\angstrom} have been calculated in the same way since that study and for our range of atmospheric parameters, the possible changes to the lines are within our uncertainties (R. Collet, private communication). More recently, studies of 3D non-LTE effects on the iron abundance have been carried out by \citet{Amarsi2016} however only for metal-poor stars ($\rm [Fe/H]<\SI{-1.5}{dex}$). The trend found is that the corrections needed decrease with increasing metallicity  \citep[see Fig.~4 in][]{Amarsi2016}. One should in general not extrapolate the results outside the parameter range studied. If however the trend were to continue towards solar metallicity, the corrections needed for our targets would probably be close to or within the internal precision on our result. 
Another necessary input is a set of atomic and molecular data, where we chose to use the \citet{Grevesse1998} values for the solar atomic abundances. 
We also tested MARCS atmosphere models, but found very little to no difference in the results with typical changes below \SI{10}{K}, \SI{0.01}{dex} and \SI{0.01}{dex} for effective temperature, surface gravity and metallicity respectively. This has previously been studied in detail by e.g. \citet{Gustafsson2008} and we will not explore it further here.

The determination of atmospheric parameters was done by requiring that: 1) [Fe/H] has no systematic dependence on the excitation potential of the FeI lines, i.e. requiring that the atmosphere is in excitation equilibrium, 2) [Fe/H] has no systematic dependence on the strength of the FeI lines and 3) the mean [Fe/H] values derived from FeI and FeII lines are consistent, i.e. requiring the atmosphere to be in ionization equilibrium. The value of [Fe/H] as a function of excitation potential is sensitive to the effective temperature, and is sensitive to the microturbulence as a function of the reduced equivalent widths of the lines ($\log (\text{EW})/\lambda$) because the microturbulence has a larger effect on stronger lines. The surface gravity is determined through its effect on the electron pressure in the stellar atmosphere with the ionization equilibrium between FeI and FeII, as the FeII lines are much more sensitive to pressure changes than the FeI lines. This is however also affected by the effective temperature and metallicity, which makes it necessary to make a number of iterations. For the NGC\,6819 and M67 targets, we also calculated a new asteroseismic $\log g$ with the newly found effective temperatures. The possible variation on the effective temperature is however low enough, that the asteroseismic $\log g$ was not significantly affected. The final results for the atmospheric parameters are presented in Table~\ref{tab:param_spec} (all targets) and Fig.~\ref{fig:comp_res} (for M67, labelled Daospec + C14).

\subsection{Metallicity of NGC\,6819}

As mentioned previously, an asteroseismic value of $\log g$ is also available for the NGC\,6819 target which allows us to use it as a calibrator for our method. When the spectroscopic analysis was done for this star with the same setup as for the M67 target, the obtained spectroscopic $\log g$ (\SI{2.52}{dex}) was in excellent agreement with the prediction from asteroseismology (\SI{2.55}{dex}, see Sec.~\ref{sec:prior}). The final [Fe/H] for NGC\,6819 was found to be $-0.02\pm$\SI{0.01}{dex} which is on the low end of the previously found values in the literature; in agreement with the result from \citet{Lee-Brown2015} ([Fe/H]=$-0.02\pm$\SI{0.02}{dex}) and slightly lower than that from \citet{Bragaglia2001} ([Fe/H]=$0.09\pm$\SI{0.03}{dex}). As can be seen from Fig.\ref{fig:comp_res} the final combination of line list and program gives results in the lower end of the range obtained for all three atmospheric parameters for the M67 target and this is also the case for the NGC\,6819 target. If we instead use the combination of ARES + C14, we get $\rm [Fe/H] = +0.11 \pm 0.01$ which is in better agreement with the high-resolution study from \citet{Bragaglia2001} who did not have asteroseismology available to constrain the surface gravity. We however also discarded this combination for NGC\,6819 because the obtained surface gravity is too high ($\log g = 2.65$) to be in agreement with asteroseismology. When doing this comparison it is important to note that within a cluster, the spread in abundances of stars can be on the order of the different results for the NGC\,6819 target we find \citep[see e.g.][for a study on abundance scatter in the Hyades]{Liu2016} and since our target is not in either of the two other NGC\,6819 studies mentioned here, a direct comparison cannot be done.

The individual abundances for all three targets are listed in Table~\ref{tab:abund}. As expected from other studies \citep{Bragaglia2001,Friel2010,Onehag2014} there is no clear sign of alpha enhancement and the abundances are mostly close to solar, with a few exceptions such as Titanium for NGC\,6819. The Magnesium and Yttrium abundances were discussed in more detail in \citet{Slumstrup2017} because it has been shown by e.g. \citet{Nissen2017} that [Y/Mg] is tightly correlated with age for solar-like stars.

\begin{table*}\small
        \centering
        \caption{\textsl{Atmospheric parameters and abundances with only internal uncertainties from spectroscopy with DAOSPEC + C14. The analysis presented in this paper shows that to a different setup, differences could be up towards $\sim$\SI{170}{\kelvin} for $T_\text{eff}$, $\sim$\SI{0.4}{dex} for $\log g$, and $\sim$\SI{0.25}{dex} for the metallicity, which should be kept in mind when comparing to other results.}}
        \begin{threeparttable}
                \begin{tabular}{l c c c c c c }  
                        \toprule
                        Target & $T_\text{eff}$ [K] & $\log g$ & $v_\text{mic}$ [km/s] & $[$Fe/H] & $n$\tnote{*} & $[\alpha$/Fe] \\
                        \midrule
                        NGC\,6819-KIC5024327 & 4695 $\pm$ 18 & 2.52 $\pm$ 0.05 & 1.19 $\pm$ 0.05 & -0.02 $\pm$ 0.01 & 97/12 & -0.02 $\pm$ 0.02\\
                        M67-EPIC211415732  & 4680 $\pm$ 18 & 2.43 $\pm$ 0.04 & 1.21 $\pm$ 0.05 & -0.03 $\pm$ 0.01 & 97/12 & 0.03 $\pm$ 0.02\\
                        NGC\,188-5085        & 4580 $\pm$ 23 & 2.51 $\pm$ 0.06 & 1.17 $\pm$ 0.06 & 0.04 $\pm$ 0.01 & 95/12 & 0.00 $\pm$ 0.02\\
                        \bottomrule
                \end{tabular}
                \footnotesize
                \begin{tablenotes}
                        \item[*] The number of FeI/FeII lines used.
                \end{tablenotes}
        \end{threeparttable}
        \label{tab:param_spec}
\end{table*}

\begin{table}\small
	\centering
	\caption{\textsl{Individual Abundances}}
	\begin{threeparttable}
		\begin{tabular}{l c c c c}
			\toprule
			 & NGC\,6819 & M67 & NGC\,188 &  $n$\tnote{*}\\
			\midrule
			$[$Na/Fe]     & 0.10  $\pm$ 0.00  & 0.18  $\pm$ 0.04  & 0.15  $\pm$ 0.03  & 2\\
			$[$Mg/Fe]     & -0.05 $\pm$ 0.03  & 0.06  $\pm$ 0.03  & 0.04  $\pm$ 0.04  & 4\\
			$[$Al/Fe]     & 0.02  $\pm$ 0.05  & 0.11  $\pm$ 0.06  & 0.14  $\pm$ 0.07  & 2\\
			$[$Si/Fe]     & 0.09  $\pm$ 0.04  & 0.14  $\pm$ 0.04  & 0.14  $\pm$ 0.04  & 14\\
			$[$Ca/Fe]     & -0.11 $\pm$ 0.06  & -0.01 $\pm$ 0.05  & -0.09 $\pm$ 0.08  & 7\\
   			$[$Ti/Fe]     & -0.24 $\pm$ 0.05  & -0.11 $\pm$ 0.03  & -0.12 $\pm$ 0.04  & 19\\
			$[$Cr/Fe]     & -0.08 $\pm$ 0.04  & -0.02 $\pm$ 0.04  & -0.03 $\pm$ 0.05  & 18\\
			$[$Ni/Fe]     & 0.00  $\pm$ 0.03  & 0.09  $\pm$ 0.03  & 0.12  $\pm$ 0.03  & 32\\
			$[$Zn/Fe]     & -0.17 $\pm$ 0.04  & -0.07 $\pm$ 0.14  & -0.11 $\pm$ 0.19  & 2\\
			$[$Y/Fe]      & 0.03  $\pm$ 0.04  & 0.07  $\pm$ 0.05  & -0.02 $\pm$ 0.06  & 3\\
			\bottomrule
		\end{tabular}
		\footnotesize
		\begin{tablenotes}
		\item[*] The number of absorption lines used in the determination. 
		\end{tablenotes}
	\end{threeparttable}
	\label{tab:abund}
\end{table}

\section{Internal uncertainties}
\label{sec:uncertainty}

In this paper we have discussed possible differences and the effect on stellar atmospheric parameters from choice of line list, atomic parameters and methods to measure the equivalent widths. 
On top of these there is the internal precision of the method itself.

When determining $T_\text{eff}$ and microturbulence, $v_\text{mic}$, the uncertainty on the final zero-slopes of [Fe/H] vs. excitation potential and [Fe/H] vs. reduced equivalent width (e.g. Fig.~\ref{fig:labvsastro}) can be easily calculated. To find the internal uncertainty on $T_\text{eff}$ and microturbulence, these parameters are varied until a $3\sigma$ difference is produced in one of the two slopes or in the difference between FeI and FeII abundances. The change in $T_\text{eff}$ or microturbulence is then divided by the highest produced difference to give the final $1\sigma$ uncertainty, which corresponds to the internal uncertainty for these parameters listed in Table ~\ref{tab:param_spec}. 

To calculate the internal $\log g$ uncertainty we can use the difference between FeI and FeII abundances determined as: $\Delta \text{[Fe/H]} / \sigma_{\Delta \text{[Fe/H]}}$ where
\begin{align}
\sigma_{\Delta \text{[Fe/H]}} = \sqrt{\left( \frac{\sigma_{\text{FeI}}}{\sqrt{n_\text{FeI}-1}} \right) ^2 + \left( \frac{\sigma_{\text{FeII}}}{\sqrt{n_\text{FeII}-1}} \right) ^2} \ .
\end{align}
Here, $\sigma_\text{FeI,II}$ is the RMS scatter on the measured abundances of each ionization stage and $n_\text{FeI,II}$ is the amount of lines used for the measurements. Once again, we vary $\log g$ until a $3\sigma$ difference is produced in one of the two slopes or in the difference between FeI and FeII abundances. 
The internal uncertainty on [Fe/H] listed in Table~\ref{tab:param_spec} corresponds to the standard error of the mean.

Another way of characterizing the internal uncertainties is to test the stability of the results. This can be done with a sensitivity analysis, which is a test of how sensitive the final metallicity is to changes in the other atmospheric parameters. This was done for the M67 target and the results are presented in Table~\ref{tab:sens}. It was only done for one target because the quality of the spectra for each star and their atmospheric parameters are so similar that the sensitivity can be assumed to be almost the same for all of them. Comparison of these results to the internal uncertainties in Table~\ref{tab:param_spec} reveals that the stability of the analysis is high, because the internal uncertainties are much lower than the the variation in the atmospheric parameters in the sensitivity analysis.
\begin{table}\small
	\centering
	\caption{\textsl{Sensitiviy of the metallicity to the other atmospheric parameters for the M67 target.}}
	\begin{threeparttable}
		\begin{tabular}{l c c c}  
			\toprule
			 & +100 K $T_\text{eff}$ & + 0.1 dex $\log g $ & +0.1 km/s $v_\text{mic}$\\
			\midrule
			$\Delta [$Fe/H] & +0.03 dex & +0.02 dex &  -0.03 dex \\
			\bottomrule
		\end{tabular}
		\footnotesize
	\end{threeparttable}
	\label{tab:sens}
\end{table}

On top of the internal uncertainties described in this section, there is significant scatter arising from using different approaches (e.g. line lists, measurements of equivalent widths etc.) as discussed in this paper. This is often not characterized, but instead an extra \SI{0.1}{dex} is simply added to the internal uncertainty on [Fe/H] and 0.1-\SI{0.2}{dex} for $\log g$, while a systematic uncertainty of about \SI{100}{\kelvin} is typically added for the effective temperature \citep[see e.g. ][]{Smalley2005,Bruntt2010,Bruntt2012}. Our analysis shows that the differences can be much larger than this, from Fig.~\ref{fig:comp_res}: $\sim$\SI{170}{\kelvin} for $T_\text{eff}$, $\sim$\SI{0.4}{dex} for $\log g$, and $\sim$\SI{0.25}{dex} for the metallicity. This should be kept in mind when comparing to other results that make use of a different combination of line list, program for measuring line strengths and atomic data.

\section{Conclusion}

We have presented a detailed spectroscopic study of three targets in three open clusters, NGC\,6819, M67 and NGC\,188, all in the same evolutionary stage and observed with the exact same instrumental setup. To our knowledge the data for the NGC\,6819 and NGC\,188 targets are of higher quality (higher resolution and higher S/N) than previously studied as described in the introduction. Along with this, we had asteroseismic data available for the two targets in NGC\,6819 and M67, which allowed us to use the M67 target as a benchmark in the analysis and the NGC\,6819 target to show the method was properly calibrated. In turn, it helped in choosing between different combinations of line lists and equivalent width measurement tools. Testing the different line lists and programs for measuring equivalent widths allowed us to characterize systematic differences within the method of a line by line spectroscopic analysis utilizing excitation and ionization equilibria in the stellar atmosphere.

The differences found on the atmospheric parameters for the M67 target with the different combinations were shown in Fig.~\ref{fig:comp_res}, where the scatter found in effective temperature is $\sim$\SI{170}{\kelvin}, $\sim$\SI{0.4}{dex} in surface gravity, and $\sim$\SI{0.25}{dex} in metallicity. The spread on the results for surface gravity is particularly high compared to the internal uncertainty obtained by each combination of line list and equivalent width measurement program. Asteroseismology allowed us to choose the most robust option for this type of targets, which was the combination of DAOSPEC and the line list based on C14. This was also the result in closets agreement with the effective temperature from photometry. 

In the current era of large spectroscopic surveys, this study highlights the possible pitfalls still existing in high precision spectroscopic analysis and why it is of crucial importance to have external constraints to calibrate the results to achieve high accuracy.

Using this setup, we also established the metallicity of NGC\,6819, an old open cluster in the \textsl{Kepler} field, to be close to solar, [Fe/H]=-0.02$\pm$\SI{0.01}{dex}. This result is on the lower end of previous determinations of the iron abundance for this cluster \citep[see e.g.][]{Bragaglia2001,Lee-Brown2015}, and it highlights the importance of carefully selecting and measuring adequate absorption lines with good atomic data for a specific set of targets.

\begin{acknowledgements}
We thank the anonymous referee for her/his careful revision of the manuscript and constructive comments that helped improve the quality of the paper. We thank Remo Collet for fruitful discussion on 3D effects. We thank Mikkel N. Lund for help with the figures. This work has made use of the VALD database, operated at Uppsala University, the Institute of Astronomy RAS in Moscow, and the University of Vienna. This research has made use of the SIMBAD database, operated at CDS, Strasbourg, France. Funding for the Stellar Astrophysics Centre is provided by The Danish National Research Foundation (Grant DNRF106). V.S.A. acknowledges support from VILLUM FONDEN (research grant 10118) and the Independent Research Fund Denmark (Research grant 7027-00096B).
\end{acknowledgements}

\bibliographystyle{aa}
\bibliography{library,moog}

\begin{thebibliography}{84}
\expandafter\ifx\csname natexlab\endcsname\relax\def\natexlab#1{#1}\fi

\bibitem[{Amarsi {et~al.}(2016)Amarsi, Lind, Asplund, Barklem, \&
  Collet}]{Amarsi2016}
Amarsi, A.~M., Lind, K., Asplund, M., Barklem, P.~S., \& Collet, R. 2016,
  \mnras, 463, 1518

\bibitem[{Asplund(2005)}]{Asplund2005}
Asplund, M. 2005, Annu. Rev. Astron. Astrophys., 43, 481

\bibitem[{Bergemann {et~al.}(2012)Bergemann, Lind, Collet, Magic, \&
  Asplund}]{Bergemann2012}
Bergemann, M., Lind, K., Collet, R., Magic, Z., \& Asplund, M. 2012, \mnras,
  427, 27

\bibitem[{Borucki {et~al.}(1997)Borucki, Koch, Dunham, \&
  Jenkins}]{Borucki1997}
Borucki, W.~J., Koch, D.~G., Dunham, E.~W., \& Jenkins, J.~M. 1997, in Planets
  Beyond Sol. Syst. Next Gener. Sp. Mission., ed. D.~R. Soderblom, 21

\bibitem[{Bragaglia {et~al.}(2001)Bragaglia, Carretta, Gratton, Tosi, Bonanno,
  Bruno, Cal{\`{i}}, Claudi, Cosentino, Desidera, Farisato, Rebeschini, \&
  Scuderi}]{Bragaglia2001}
Bragaglia, A., Carretta, E., Gratton, R., {et~al.} 2001, \aj, 121, 327

\bibitem[{Brogaard {et~al.}(2017)Brogaard, Hansen, Miglio, Slumstrup, Frandsen,
  Jessen-Hansen, Lund, Bossini, Thygesen, Davies, Chaplin, Arentoft, Bruntt,
  Grundahl, \& Handberg}]{Brogaard2017}
Brogaard, K., Hansen, C.~J., Miglio, A., {et~al.} 2017, \mnras, 000, 1

\bibitem[{Brogaard {et~al.}(2012)Brogaard, VandenBerg, Bruntt, Grundahl,
  Frandsen, Bedin, Milone, Dotter, Feiden, Stetson, Sandquist, Miglio, Stello,
  \& Jessen-Hansen}]{Brogaard2012}
Brogaard, K., VandenBerg, D.~A., Bruntt, H., {et~al.} 2012, \aap, 543, A106

\bibitem[{Brown {et~al.}(1991)Brown, Gilliland, Noyes, \& Ramsey}]{Brown1991}
Brown, T.~M., Gilliland, R.~L., Noyes, R.~W., \& Ramsey, L.~W. 1991, \apj, 368,
  599

\bibitem[{Bruntt {et~al.}(2012)Bruntt, Basu, Smalley, Chaplin, Verner, Bedding,
  Catala, Gazzano, Molenda-Zakowicz, Thygesen, Uytterhoeven, Hekker, Huber,
  Karoff, Mathur, Mosser, Appourchaux, Campante, Elsworth, Garc{\'{i}}a,
  Handberg, Metcalfe, Quirion, R{\'{e}}gulo, Roxburgh, Stello,
  Christensen-Dalsgaard, Kawaler, Kjeldsen, Morris, Quintana, \&
  Sanderfer}]{Bruntt2012}
Bruntt, H., Basu, S., Smalley, B., {et~al.} 2012, \mnras, 423, 122

\bibitem[{Bruntt {et~al.}(2010)Bruntt, Bedding, Quirion, {Lo Curto}, Carrier,
  Smalley, Dall, Arentoft, Bazot, \& Butler}]{Bruntt2010}
Bruntt, H., Bedding, T.~R., Quirion, P.~O., {et~al.} 2010, \mnras, 405, 1907

\bibitem[{Bruntt {et~al.}(2011)Bruntt, Frandsen, \& Thygesen}]{Bruntt2011}
Bruntt, H., Frandsen, S., \& Thygesen, A.~O. 2011, \aap, 528, A121

\bibitem[{Carraro {et~al.}(2014)Carraro, Villanova, Monaco, Beccari, Ahumada,
  \& Boffin}]{Carraro2014a}
Carraro, G., Villanova, S., Monaco, L., {et~al.} 2014, \aap, 562, A39

\bibitem[{Casagrande {et~al.}(2014)Casagrande, Aguirre, Stello, Huber,
  Serenelli, Cassisi, Dotter, Milone, Hodgkin, Marino, Lund, Pietrinferni,
  Asplund, Feltzing, Flynn, Grundahl, Nissen, Sch{\"{o}}nrich, Schlesinger, \&
  Wang}]{Casagrande2014}
Casagrande, L., Aguirre, V.~S., Stello, D., {et~al.} 2014, \apj, 787, 110

\bibitem[{Casagrande {et~al.}(2016)Casagrande, {Silva Aguirre}, Schlesinger,
  Stello, Huber, Serenelli, Sch{\"{o}}nrich, Cassisi, Pietrinferni, Hodgkin,
  Milone, Feltzing, \& Asplund}]{Casagrande2016}
Casagrande, L., {Silva Aguirre}, V., Schlesinger, K.~J., {et~al.} 2016, \mnras,
  455, 987

\bibitem[{Casamiquela {et~al.}(2017)Casamiquela, Carrera, Blanco-Cuaresma,
  Jordi, Balaguer-N{\'{u}}{\~{n}}ez, Pancino, Anders, Chiappini,
  D{\'{i}}az-P{\'{e}}rez, Aguado, Aparicio, Garcia-Dias, Heiter,
  Mart{\'{i}}nez-V{\'{a}}zquez, Murabito, \& del Pino}]{Casamiquela2017}
Casamiquela, L., Carrera, R., Blanco-Cuaresma, S., {et~al.} 2017, \mnras, 470,
  4363

\bibitem[{Castelli \& Kurucz(2004)}]{Castelli2004}
Castelli, F. \& Kurucz, R.~L. 2004, ArXiv Astrophys. e-prints
  [\eprint[arXiv]{0405087}]

\bibitem[{Collet {et~al.}(2007)Collet, Asplund, \& Trampedach}]{Collet2007}
Collet, R., Asplund, M., \& Trampedach, R. 2007, \aap, 469, 687

\bibitem[{Corsaro {et~al.}(2012)Corsaro, Stello, Huber, Bedding, Bonanno,
  Brogaard, Kallinger, Benomar, White, Mosser, Basu, Chaplin,
  Christensen-Dalsgaard, Elsworth, Garc{\'{i}}a, Hekker, Kjeldsen, Mathur,
  Meibom, Hall, Ibrahim, \& Klaus}]{Corsaro2012}
Corsaro, E., Stello, D., Huber, D., {et~al.} 2012, \apj, 757, 190

\bibitem[{{De Silva} {et~al.}(2015){De Silva}, Freeman, Bland-Hawthorn,
  Martell, {Wylie de Boer}, Asplund, Keller, Sharma, Zucker, Zwitter, Anguiano,
  Bacigalupo, Bayliss, Beavis, Bergemann, Campbell, Cannon, Carollo,
  Casagrande, Casey, {Da Costa}, D'Orazi, Dotter, Duong, Heger, Ireland, Kafle,
  Kos, Lattanzio, Lewis, Lin, Lind, Munari, Nataf, O'Toole, Parker, Reid,
  Schlesinger, Sheinis, Simpson, Stello, Ting, Traven, Watson, Wittenmyer,
  Yong, \& {\v{Z}}erjal}]{DeSilva2015}
{De Silva}, G.~M., Freeman, K.~C., Bland-Hawthorn, J., {et~al.} 2015, \mnras,
  449, 2604

\bibitem[{Doyle {et~al.}(2017)Doyle, Smalley, Faedi, Pollacco, \& {G{\'{o}}mez
  Maqueo Chew}}]{Doyle2017}
Doyle, A.~P., Smalley, B., Faedi, F., Pollacco, D., \& {G{\'{o}}mez Maqueo
  Chew}, Y. 2017, \mnras, 469, 4850

\bibitem[{Friel {et~al.}(2010)Friel, Jacobson, \& Pilachowski}]{Friel2010}
Friel, E.~D., Jacobson, H.~R., \& Pilachowski, C.~A. 2010, \aj, 139, 1942

\bibitem[{{Garc{\'{i}}a P{\'{e}}rez} {et~al.}(2016){Garc{\'{i}}a P{\'{e}}rez},
  {Allende Prieto}, Holtzman, Shetrone, M{\'{e}}sz{\'{a}}ros, Bizyaev, Carrera,
  Cunha, Garc{\'{i}}a-Hern{\'{a}}ndez, Johnson, Majewski, Nidever, Schiavon,
  Shane, Smith, Sobeck, Troup, Zamora, Weinberg, Bovy, Eisenstein, Feuillet,
  Frinchaboy, Hayden, Hearty, Nguyen, O'Connell, Pinsonneault, Wilson, \&
  Zasowski}]{Perez2015}
{Garc{\'{i}}a P{\'{e}}rez}, A.~E., {Allende Prieto}, C., Holtzman, J.~A.,
  {et~al.} 2016, \aj, 151, 144

\bibitem[{Gilmore {et~al.}(2012)Gilmore, Randich, Asplund, Binney, Bonifacio,
  Drew, Feltzing, Ferguson, Jeffries, Micela, Negueruela, Prusti, Rix,
  Vallenari, Alfaro, Allende-Prieto, Babusiaux, Bensby, Blomme, Bragaglia,
  Flaccomio, Fran{\c{c}}ois, Irwin, Koposov, Korn, Lanzafame, Pancino, Paunzen,
  Recio-Blanco, Sacco, Smiljanic, {Van Eck}, Walton, Aden, Aerts, Affer,
  Alcala, Altavilla, Alves, Antoja, Arenou, Argiroffi, {Asensio Ramos},
  Bailer-Jones, Balaguer-Nunez, Bayo, Barbuy, Barisevicius, {Barrado y
  Navascues}, Battistini, {Bellas Velidis}, Bellazzini, Belokurov, Bergemann,
  Bertelli, Biazzo, Bienayme, Bland-Hawthorn, Boeche, Bonito, Boudreault,
  Bouvier, Brandao, Brown, de~Bruijne, Burleigh, Caballero, Caffau, Calura,
  Capuzzo-Dolcetta, Caramazza, Carraro, Casagrande, Casewell, Chapman,
  Chiappini, Chorniy, Christlieb, Cignoni, Cocozza, Colless, Collet, Collins,
  Correnti, Covino, Crnojevic, Cropper, Cunha, Damiani, David, Delgado, Duffau,
  Edvardsson, Eldridge, Enke, Eriksson, Evans, Eyer, Famaey, Fellhauer,
  Ferreras, Figueras, Fiorentino, Flynn, Folha, Franciosini, Frasca, Freeman,
  Fremat, Friel, Gaensicke, Gameiro, Garzon, Geier, Geisler, Gerhard, Gibson,
  Gomboc, Gomez, Gonzalez-Fernandez, {Gonzalez Hernandez}, Gosset, Grebel,
  Greimel, Groenewegen, Grundahl, Guarcello, Gustafsson, Hadrava,
  Hatzidimitriou, Hambly, Hammersley, Hansen, Haywood, Heber, Heiter, Held,
  Helmi, Hensler, Herrero, Hill, Hodgkin, Huelamo, Huxor, Ibata, Jackson,
  de~Jong, Jonker, Jordan, Jordi, Jorissen, Katz, Kawata, Keller, Kharchenko,
  Klement, Klutsch, Knude, Koch, Kochukhov, Kontizas, Koubsky, Lallement,
  de~Laverny, van Leeuwen, Lemasle, Lewis, Lind, Lindstrom, Lobel, {Lopez
  Santiago}, Lucas, Ludwig, Lueftinger, Magrini, {Maiz Apellaniz}, Maldonado,
  Marconi, Marino, Martayan, Martinez-Valpuesta, Matijevic, McMahon, Messina,
  Meyer, Miglio, Mikolaitis, Minchev, Minniti, Moitinho, Momany, Monaco,
  Montalto, Monteiro, Monier, Montes, Mora, Moraux, Morel, Mowlavi,
  Mucciarelli, Munari, Napiwotzki, Nardetto, Naylor, Naze, Nelemans, Okamoto,
  Ortolani, Pace, Palla, Palous, Parker, Penarrubia, Pillitteri, Piotto,
  Posbic, Prisinzano, Puzeras, Quirrenbach, Ragaini, Read, Read, Reyle, {De
  Ridder}, Robichon, Robin, Roeser, Romano, Royer, Ruchti, Ruzicka, Ryan, Ryde,
  Santos, {Sanz Forcada}, {Sarro Baro}, Sbordone, Schilbach, Schmeja, Schnurr,
  Schoenrich, Scholz, Seabroke, Sharma, {De Silva}, Smith, Solano, Sordo,
  Soubiran, Sousa, Spagna, Steffen, Steinmetz, Stelzer, Stempels, Tabernero,
  Tautvaisiene, Thevenin, Torra, Tosi, Tolstoy, Turon, Walker, Wambsganss,
  Worley, Venn, Vink, Wyse, Zaggia, Zeilinger, Zoccali, Zorec, Zucker, Zwitter,
  \& {Gaia-ESO Survey Team}}]{Gilmore2012}
Gilmore, G., Randich, S., Asplund, M., {et~al.} 2012, The Messenger, 147, 25

\bibitem[{Gray \& Corbally(1994)}]{Gray1994}
Gray, R.~O. \& Corbally, C.~J. 1994, \aj, 107, 742

\bibitem[{Grevesse \& Sauval(1998)}]{Grevesse1998}
Grevesse, N. \& Sauval, A.~J. 1998, Sp. Sci. Rev., 85, 161

\bibitem[{Gustafsson {et~al.}(2008)Gustafsson, Edvardsson, Eriksson,
  J{\o}rgensen, Nordlund, \& Plez}]{Gustafsson2008}
Gustafsson, B., Edvardsson, B., Eriksson, K., {et~al.} 2008, \aap, 486, 951

\bibitem[{Handberg {et~al.}(2017)Handberg, Brogaard, Miglio, Bossini, Elsworth,
  Slumstrup, Davies, \& Chaplin}]{Handberg2017}
Handberg, R., Brogaard, K., Miglio, A., {et~al.} 2017, \mnras, 472, 979

\bibitem[{Hawkins {et~al.}(2016)Hawkins, Masseron, Jofr{\'{e}}, Gilmore,
  Elsworth, \& Hekker}]{Hawkins2016}
Hawkins, K., Masseron, T., Jofr{\'{e}}, P., {et~al.} 2016, \aap, 594, A43

\bibitem[{Heiter {et~al.}(2015)Heiter, Jofr{\'{e}}, Gustafsson, Korn, Soubiran,
  \& Th{\'{e}}venin}]{Heiter2015}
Heiter, U., Jofr{\'{e}}, P., Gustafsson, B., {et~al.} 2015, \aap, 49, 1

\bibitem[{Hinkel {et~al.}(2016)Hinkel, Young, Pagano, Desch, Anbar, Adibekyan,
  Blanco-Cuaresma, Carlberg, Mena, Liu, Nordlander, Sousa, Korn, Gruyters,
  Heiter, Jofre, Santos, \& Soubiran}]{Hinkel2016}
Hinkel, N.~R., Young, P.~A., Pagano, M.~D., {et~al.} 2016, Astrophys. J. Suppl.
  Ser., 226, 1

\bibitem[{Hole {et~al.}(2009)Hole, Geller, Mathieu, Platais, Meibom, \&
  Latham}]{Hole2009}
Hole, K.~T., Geller, A.~M., Mathieu, R.~D., {et~al.} 2009, \aj, 138, 159

\bibitem[{Howell {et~al.}(2014)Howell, Sobeck, Haas, Still, Barclay, Mullally,
  Troeltzsch, Aigrain, Bryson, Caldwell, Chaplin, Cochran, Huber, Marcy,
  Miglio, Najita, Smith, Twicken, \& Fortney}]{Howell2014}
Howell, S.~B., Sobeck, C., Haas, M., {et~al.} 2014, \pasp, 126, 398

\bibitem[{Huber {et~al.}(2012)Huber, Ireland, Bedding, Brand{\~{a}}o, Piau,
  Maestro, White, Bruntt, Casagrande, Molenda-{\.{Z}}akowicz, Aguirre, Sousa,
  Barclay, Burke, Chaplin, Christensen-Dalsgaard, Cunha, {De Ridder},
  Farrington, Frasca, Garc{\'{i}}a, Gilliland, Goldfinger, Hekker, Kawaler,
  Kjeldsen, McAlister, Metcalfe, Miglio, Monteiro, Pinsonneault, Schaefer,
  Stello, Stumpe, Sturmann, Sturmann, ten Brummelaar, Thompson, Turner, \&
  Uytterhoeven}]{Huber2012}
Huber, D., Ireland, M.~J., Bedding, T.~R., {et~al.} 2012, \apj, 760, 32

\bibitem[{Huber {et~al.}(2017)Huber, Zinn, Bojsen-Hansen, Pinsonneault,
  Sahlholdt, Serenelli, {Silva Aguirre}, Stassun, Stello, Tayar, Bastien,
  Bedding, Buchhave, Chaplin, Davies, Garcia, Latham, Mathur, Mosser, \&
  Sharma}]{Huber2017}
Huber, D., Zinn, J., Bojsen-Hansen, M., {et~al.} 2017, \apj, 844, 102

\bibitem[{Jofr{\'{e}} {et~al.}(2018)Jofr{\'{e}}, Heiter, \&
  Soubiran}]{Jofre2018}
Jofr{\'{e}}, P., Heiter, U., \& Soubiran, C. 2018 [\eprint[arXiv]{1811.08041}]

\bibitem[{Jofr{\'{e}} {et~al.}(2014)Jofr{\'{e}}, Heiter, Soubiran,
  Blanco-Cuaresma, Worley, Pancino, Cantat-Gaudin, Magrini, Bergemann,
  {Gonz{\'{a}}lez Hern{\'{a}}ndez}, Hill, Lardo, de~Laverny, Lind, Masseron,
  Montes, Mucciarelli, Nordlander, {Recio Blanco}, Sobeck, Sordo, Sousa,
  Tabernero, Vallenari, \& {Van Eck}}]{Jofre2014}
Jofr{\'{e}}, P., Heiter, U., Soubiran, C., {et~al.} 2014, \aap, 564, A133

\bibitem[{Kang \& Lee(2012)}]{Kang2012}
Kang, W. \& Lee, S.-G. 2012, \mnras, 425, 3162

\bibitem[{Kjeldsen \& Bedding(1995)}]{Kjeldsen1995}
Kjeldsen, H. \& Bedding, T.~R. 1995, \aap, 293, 87

\bibitem[{Lebzelter {et~al.}(2012)Lebzelter, Heiter, Abia, Eriksson, Ireland,
  Neilson, Nowotny, Maldonado, Merle, Peterson, Plez, Short, Wahlgren, Worley,
  Aringer, Bladh, de~Laverny, Goswami, Mora, Norris, Recio-Blanco, Scholz,
  Th{\'{e}}venin, Tsuji, Kordopatis, Montesinos, \& Wing}]{Lebzelter2012}
Lebzelter, T., Heiter, U., Abia, C., {et~al.} 2012, \aap, 547, A108

\bibitem[{Lee-Brown {et~al.}(2015)Lee-Brown, Anthony-Twarog, Deliyannis, Rich,
  \& Twarog}]{Lee-Brown2015}
Lee-Brown, D.~B., Anthony-Twarog, B.~J., Deliyannis, C.~P., Rich, E., \&
  Twarog, B.~A. 2015, \aj, 149, 121

\bibitem[{Lind {et~al.}(2012)Lind, Bergemann, \& Asplund}]{Lind2012}
Lind, K., Bergemann, M., \& Asplund, M. 2012, \mnras, 427, 50

\bibitem[{Liu {et~al.}(2016)Liu, Yong, Asplund, Ram{\'{i}}rez, \&
  Mel{\'{e}}ndez}]{Liu2016}
Liu, F., Yong, D., Asplund, M., Ram{\'{i}}rez, I., \& Mel{\'{e}}ndez, J. 2016,
  \mnras, 457, 3934

\bibitem[{Majewski {et~al.}(2017)Majewski, Schiavon, Frinchaboy, {Allende
  Prieto}, Barkhouser, Bizyaev, Blank, Brunner, Burton, Carrera, Chojnowski,
  Cunha, Epstein, Fitzgerald, {Garc{\'{i}}a P{\'{e}}rez}, Hearty, Henderson,
  Holtzman, Johnson, Lam, Lawler, Maseman, M{\'{e}}sz{\'{a}}ros, Nelson,
  Nguyen, Nidever, Pinsonneault, Shetrone, Smee, Smith, Stolberg, Skrutskie,
  Walker, Wilson, Zasowski, Anders, Basu, Beland, Blanton, Bovy, Brownstein,
  Carlberg, Chaplin, Chiappini, Eisenstein, Elsworth, Feuillet, Fleming,
  Galbraith-Frew, Garc{\'{i}}a, Garc{\'{i}}a-Hern{\'{a}}ndez, Gillespie,
  Girardi, Gunn, Hasselquist, Hayden, Hekker, Ivans, Kinemuchi, Klaene,
  Mahadevan, Mathur, Mosser, Muna, Munn, Nichol, O'Connell, Parejko, Robin,
  Rocha-Pinto, Schultheis, Serenelli, Shane, {Silva Aguirre}, Sobeck, Thompson,
  Troup, Weinberg, \& Zamora}]{Majewski2015}
Majewski, S.~R., Schiavon, R.~P., Frinchaboy, P.~M., {et~al.} 2017, \aj, 154,
  94

\bibitem[{Mashonkina {et~al.}(2011)Mashonkina, Gehren, Shi, Korn, \&
  Grupp}]{Mashonkina2011}
Mashonkina, L.~I., Gehren, T., Shi, J.-R., Korn, A.~J., \& Grupp, F. 2011,
  \aap, 528, A87

\bibitem[{Meibom {et~al.}(2009)Meibom, Grundahl, Clausen, Mathieu, Frandsen,
  Pigulski, Narwid, Steslicki, \& Lefever}]{Meibom2009}
Meibom, S., Grundahl, F., Clausen, J.~V., {et~al.} 2009, \aj, 137, 5086

\bibitem[{Miglio {et~al.}(2012)Miglio, Brogaard, Stello, Chaplin, D'Antona,
  Montalb{\'{a}}n, Basu, Bressan, Grundahl, Pinsonneault, Serenelli, Elsworth,
  Hekker, Kallinger, Mosser, Ventura, Bonanno, Noels, {Silva Aguirre}, Szabo,
  Li, McCauliff, Middour, \& Kjeldsen}]{Miglio2012}
Miglio, A., Brogaard, K., Stello, D., {et~al.} 2012, \mnras, 419, 2077

\bibitem[{Miglio {et~al.}(2016)Miglio, Chaplin, Brogaard, Lund, Mosser, Davies,
  Handberg, Milone, Marino, Bossini, Elsworth, Grundah, Arentoft, Bedin,
  Campante, Jessen-Hansen, Jones, Kuszlewicz, Malavolta, Nascimbeni, \&
  Sandquist}]{Miglio2016}
Miglio, A., Chaplin, W.~J., Brogaard, K., {et~al.} 2016, \mnras, 461, 760

\bibitem[{Miglio {et~al.}(2013)Miglio, Chiappini, Morel, Barbieri, Chaplin,
  Girardi, Montalb'an, Valentini, Mosser, Baudin, Casagrande, Fossati, {Silva
  Aguirre}, \& Baglin}]{Miglio2013}
Miglio, A., Chiappini, C., Morel, T., {et~al.} 2013, \mnras, 429, 423

\bibitem[{Nissen(2015)}]{Nissen2015}
Nissen, P.~E. 2015, \aap, 579, 52

\bibitem[{Nissen {et~al.}(1997)Nissen, H{\o}g, \& Schuster}]{Nissen1997}
Nissen, P.~E., H{\o}g, E., \& Schuster, W.~J. 1997, ESA Spec. Publ., 225

\bibitem[{Nissen {et~al.}(2017)Nissen, {Silva Aguirre}, Christensen-Dalsgaard,
  Collet, Grundahl, \& Slumstrup}]{Nissen2017}
Nissen, P.~E., {Silva Aguirre}, V., Christensen-Dalsgaard, J., {et~al.} 2017,
  \aap, 608, A112

\bibitem[{{\"{O}}nehag {et~al.}(2014){\"{O}}nehag, Gustafsson, \&
  Korn}]{Onehag2014}
{\"{O}}nehag, A., Gustafsson, B., \& Korn, A.~J. 2014, \aap, 562, A102

\bibitem[{Pinsonneault {et~al.}(2014)Pinsonneault, Elsworth, Epstein, Hekker,
  M{\'{e}}sz{\'{a}}ros, Chaplin, Johnson, Garc{\'{i}}a, Holtzman, Mathur,
  {Garc{\'{i}}a P{\'{e}}rez}, {Silva Aguirre}, Girardi, Basu, Shetrone, Stello,
  {Allende Prieto}, An, Beck, Beers, Bizyaev, Bloemen, Bovy, Cunha, {De
  Ridder}, Frinchaboy, Garc{\'{i}}a-Hern{\'{a}}ndez, Gilliland, Harding,
  Hearty, Huber, Ivans, Kallinger, Majewski, Metcalfe, Miglio, Mosser, Muna,
  Nidever, Schneider, Serenelli, Smith, Tayar, Zamora, \&
  Zasowski}]{Pinsonneault2015}
Pinsonneault, M.~H., Elsworth, Y., Epstein, C., {et~al.} 2014, Astrophys. J.
  Suppl. Ser., 215, 19

\bibitem[{Pinsonneault {et~al.}(2018)Pinsonneault, Elsworth, Tayar, Serenelli,
  Stello, Zinn, Mathur, Garc{\'{i}}a, Johnson, Hekker, Huber, Kallinger,
  M{\'{e}}sz{\'{a}}ros, Mosser, Stassun, Girardi, Rodrigues, {Silva Aguirre},
  An, Basu, Chaplin, Corsaro, Cunha, Garc{\'{i}}a-Hern{\'{a}}ndez, Holtzman,
  J{\"{o}}nsson, Shetrone, Smith, Sobeck, Stringfellow, Zamora, Beers,
  Fern{\'{a}}ndez-Trincado, Frinchaboy, Hearty, \&
  Nitschelm}]{Pinsonneault2018}
Pinsonneault, M.~H., Elsworth, Y.~P., Tayar, J., {et~al.} 2018
  [\eprint[arXiv]{1804.09983}]

\bibitem[{Piskunov {et~al.}(1995)Piskunov, Kupka, Ryabchikova, Weiss, \&
  Jeffery}]{Piskunov1995}
Piskunov, N.~E., Kupka, F., Ryabchikova, T.~a., Weiss, W.~W., \& Jeffery, C.~S.
  1995, Astron. Astrophys. Suppl., 112, 525

\bibitem[{Ram{\'{i}}rez \& Mel{\'{e}}ndez(2005)}]{Ramirez2005}
Ram{\'{i}}rez, I. \& Mel{\'{e}}ndez, J. 2005, \apj, 626, 465

\bibitem[{Ram{\'{i}}rez {et~al.}(2014)Ram{\'{i}}rez, Mel{\'{e}}ndez, Bean,
  Asplund, Bedell, Monroe, Casagrande, Schirbel, Dreizler, Teske, {Tucci Maia},
  Alves-Brito, \& Baumann}]{Ramirez2014b}
Ram{\'{i}}rez, I., Mel{\'{e}}ndez, J., Bean, J., {et~al.} 2014, \aap, 572, A48

\bibitem[{Randich {et~al.}(2013)Randich, Gilmore, \& Consortium}]{Randich2013}
Randich, S., Gilmore, G., \& Consortium, G.-E. 2013, The Messenger, 154, 47

\bibitem[{Randich {et~al.}(2003)Randich, Sestito, \& Pallavicini}]{Randich2003}
Randich, S., Sestito, P., \& Pallavicini, R. 2003, \aap, 399, 133

\bibitem[{Rosvick \& VandenBerg(1998)}]{Rosvick1998}
Rosvick, J.~M. \& VandenBerg, D.~A. 1998, \aj, 115, 1516

\bibitem[{Sahlholdt {et~al.}(2018)Sahlholdt, {Silva Aguirre}, Casagrande,
  Mosumgaard, \& Bojsen-Hansen}]{Sahlholdt2018}
Sahlholdt, C.~L., {Silva Aguirre}, V., Casagrande, L., Mosumgaard, J.~R., \&
  Bojsen-Hansen, M. 2018, \mnras, 476, 1931

\bibitem[{Scott {et~al.}(2015)Scott, Asplund, Grevesse, Bergemann, \& {Jacques
  Sauval}}]{Scott2014}
Scott, P., Asplund, M., Grevesse, N., Bergemann, M., \& {Jacques Sauval}, A.
  2015, \aap, 573, A26

\bibitem[{{Silva Aguirre} {et~al.}(2018){Silva Aguirre}, Bojsen-Hansen,
  Slumstrup, Casagrande, Kawata, Ciuc{\'{a}}, Handberg, Lund, Mosumgaard,
  Huber, Johnson, Pinsonneault, Serenelli, Stello, Tayar, Bird, Cassisi, Hon,
  Martig, Nissen, Rix, Sch{\"{o}}nrich, Sahlholdt, Trick, \&
  Yu}]{SilvaAguirre2018}
{Silva Aguirre}, V., Bojsen-Hansen, M., Slumstrup, D., {et~al.} 2018, \mnras,
  5500, 5487

\bibitem[{{Silva Aguirre} {et~al.}(2012){Silva Aguirre}, Casagrande, Basu,
  Campante, Chaplin, Huber, Miglio, Serenelli, Ballot, Bedding,
  Christensen-Dalsgaard, Creevey, Elsworth, Garc{\'{i}}a, Gilliland, Hekker,
  Kjeldsen, Mathur, Metcalfe, Monteiro, Mosser, Pinsonneault, Stello, Weiss,
  Tenenbaum, Twicken, \& Uddin}]{SilvaAguirre2012}
{Silva Aguirre}, V., Casagrande, L., Basu, S., {et~al.} 2012, \apj, 757, 99

\bibitem[{{Silva Aguirre} {et~al.}(2016){Silva Aguirre}, Lund, Antia, Ball,
  Basu, Christensen-Dalsgaard, Lebreton, Reese, Verma, Casagrande, Justesen,
  Mosumgaard, Chaplin, Bedding, Davies, Handberg, Houdek, Huber, Kjeldsen,
  Latham, White, Coelho, Miglio, \& Rendle}]{SilvaAguirre2017}
{Silva Aguirre}, V., Lund, M.~N., Antia, H.~M., {et~al.} 2016, \apj, 835, 1

\bibitem[{Skrutskie {et~al.}(2006)Skrutskie, Cutri, Stiening, Weinberg,
  Schneider, Carpenter, Beichman, Capps, Chester, Elias, Huchra, Liebert,
  Lonsdale, Monet, Price, Seitzer, Jarrett, Kirkpatrick, Gizis, Howard, Evans,
  Fowler, Fullmer, Hurt, Light, Kopan, Marsh, McCallon, Tam, {Van Dyk}, \&
  Wheelock}]{Skrutskie2006}
Skrutskie, M.~F., Cutri, R.~M., Stiening, R., {et~al.} 2006, \aj, 131, 1163

\bibitem[{Slumstrup {et~al.}(2017)Slumstrup, Grundahl, Brogaard, Thygesen,
  Nissen, Jessen-Hansen, {Van Eylen}, \& Pedersen}]{Slumstrup2017}
Slumstrup, D., Grundahl, F., Brogaard, K., {et~al.} 2017, \aap, 604, L8

\bibitem[{Smalley(2005)}]{Smalley2005}
Smalley, B. 2005, \memsai, 8, 130

\bibitem[{Smiljanic {et~al.}(2014)Smiljanic, Korn, Bergemann, Frasca, Magrini,
  Masseron, Pancino, Ruchti, {San Roman}, Sbordone, Sousa, Tabernero,
  Tautvai{\v{s}}ienė, Valentini, Weber, Worley, Adibekyan, {Allende Prieto},
  Barisevi{\v{c}}ius, Biazzo, Blanco-Cuaresma, Bonifacio, Bragaglia, Caffau,
  Cantat-Gaudin, Chorniy, de~Laverny, Delgado-Mena, Donati, Duffau,
  Franciosini, Friel, Geisler, {Gonz{\'{a}}lez Hern{\'{a}}ndez}, Gruyters,
  Guiglion, Hansen, Heiter, Hill, Jacobson, Jofre, J{\"{o}}nsson, Lanzafame,
  Lardo, Ludwig, Maiorca, Mikolaitis, Montes, Morel, Mucciarelli, Mu{\~{n}}oz,
  Nordlander, Pasquini, Puzeras, Recio-Blanco, Ryde, Sacco, Santos, Serenelli,
  Sordo, Soubiran, Spina, Steffen, Vallenari, {Van Eck}, Villanova, Gilmore,
  Randich, Asplund, Binney, Drew, Feltzing, Ferguson, Jeffries, Micela,
  Negueruela, Prusti, Rix, Alfaro, Babusiaux, Bensby, Blomme, Flaccomio,
  Fran{\c{c}}ois, Irwin, Koposov, Walton, Bayo, Carraro, Costado, Damiani,
  Edvardsson, Hourihane, Jackson, Lewis, Lind, Marconi, Martayan, Monaco,
  Morbidelli, Prisinzano, \& Zaggia}]{Smiljanic2014}
Smiljanic, R., Korn, A.~J., Bergemann, M., {et~al.} 2014, \aap, 570, A122

\bibitem[{Sneden(1973)}]{Sneden1973}
Sneden, C. 1973, PhD thesis, THE UNIVERSITY OF TEXAS AT AUSTIN.

\bibitem[{Sousa {et~al.}(2007)Sousa, Santos, Israelian, Mayor, \&
  Monteiro}]{Sousa2007}
Sousa, S.~G., Santos, N.~C., Israelian, G., Mayor, M., \& Monteiro, M. J. P.
  F.~G. 2007, \aap, 469, 783

\bibitem[{Sousa {et~al.}(2008)Sousa, Santos, Mayor, Udry, Casagrande,
  Israelian, Pepe, Queloz, \& Monteiro}]{Sousa2008}
Sousa, S.~G., Santos, N.~C., Mayor, M., {et~al.} 2008, \aap, 487, 373

\bibitem[{Steinmetz {et~al.}(2006)Steinmetz, Zwitter, Siebert, Watson, Freeman,
  Munari, Campbell, Williams, Seabroke, Wyse, Parker, Bienaym{\'{e}}, Roeser,
  Gibson, Gilmore, Grebel, Helmi, Navarro, Burton, Cass, Dawe, Fiegert,
  Hartley, Russell, Saunders, Enke, Bailin, Binney, Bland-Hawthorn, Boeche,
  Dehnen, Eisenstein, Evans, Fiorucci, Fulbright, Gerhard, Jauregi, Kelz,
  Mijovi{\'{c}}, Minchev, Parmentier, Pe{\~{n}}arrubia, Quillen, Read, Ruchti,
  Scholz, Siviero, Smith, Sordo, Veltz, Vidrih, von Berlepsch, Boyle, \&
  Schilbach}]{Steinmetz2006}
Steinmetz, M., Zwitter, T., Siebert, A., {et~al.} 2006, \aj, 132, 1645

\bibitem[{Stello {et~al.}(2016)Stello, Vanderburg, Casagrande, Gilliland,
  {Silva Aguirre}, Sandquist, Leiner, Mathieu, Soderblom, {Silva Aguirre},
  Sandquist, Leiner, Mathieu, \& Soderblom}]{Stello2016}
Stello, D., Vanderburg, A., Casagrande, L., {et~al.} 2016, \apj, 832, 133

\bibitem[{Stetson {et~al.}(2004)Stetson, Mcclure, \& Vandenberg}]{Stetson2004}
Stetson, P.~B., Mcclure, R.~D., \& Vandenberg, D.~A. 2004, 1012

\bibitem[{Stetson \& Pancino(2008)}]{Stetson2008}
Stetson, P.~B. \& Pancino, E. 2008, \pasp, 120, 1332

\bibitem[{Taylor(2007)}]{Taylor2007}
Taylor, B.~J. 2007, \aj, 133, 370

\bibitem[{Thygesen {et~al.}(2012)Thygesen, Frandsen, Bruntt, Kallinger,
  Andersen, Elsworth, Hekker, Karoff, Stello, Brogaard, Burke, Caldwell, \&
  Christiansen}]{Thygesen2012a}
Thygesen, A.~O., Frandsen, S., Bruntt, H., {et~al.} 2012, \aap, 160, 1

\bibitem[{Torres(2010)}]{Torres2010}
Torres, G. 2010, \aj, 140, 1158

\bibitem[{Tsantaki {et~al.}(2013)Tsantaki, Sousa, Adibekyan, Santos, Mortier,
  \& Israelian}]{Tsantaki2013}
Tsantaki, M., Sousa, S.~G., Adibekyan, V.~Z., {et~al.} 2013, \aap, 555, A150

\bibitem[{Valenti \& Piskunov(1996)}]{Valenti1996}
Valenti, J.~A. \& Piskunov, N. 1996, Astron. Astrophys. Suppl. Ser., 118, 595

\bibitem[{Valentini {et~al.}(2016)Valentini, Chiappini, Miglio,
  Montalb{\'{a}}n, Rodrigues, Mosser, Anders, CoRot, \& Ges}]{Valentini2016}
Valentini, M., Chiappini, C., Miglio, A., {et~al.} 2016, Astron. Nachrichten,
  337, 970

\bibitem[{White {et~al.}(2013)White, Huber, Maestro, Bedding, Ireland, Baron,
  Boyajian, Che, Monnier, Pope, Roettenbacher, Stello, Tuthill, Farrington,
  Goldfinger, McAlister, Schaefer, Sturmann, Sturmann, ten Brummelaar, \&
  Turner}]{White2013}
White, T.~R., Huber, D., Maestro, V., {et~al.} 2013, \mnras, 433, 1262

\bibitem[{Yadav {et~al.}(2008)Yadav, Bedin, Piotto, Anderson, Cassisi,
  Villanova, Platais, Pasquini, Momany, \& Sagar}]{Yadav2008}
Yadav, R. K.~S., Bedin, L.~R., Piotto, G., {et~al.} 2008, \aap, 484, 609

\end{thebibliography}

\appendix

\onecolumn

\section{Line list}

\begin{longtable}{c c c c c c c c c c}
    \label{tab:cluster}\\
            \toprule
            Wavelength & Element & Exc. Pot. & $\log gf$ & \multicolumn{3}{c}{EW [\si{\milli\angstrom}]}\\
            $[\si{\angstrom}]$ & & [eV] & & NGC\,6819-KIC5024327 & M67-EPIC211415732 & NGC\,188-085\\
            \midrule
     \endhead
     		\bottomrule
     \endfoot
            5052.1350  &  6.0   &  7.685  &  -1.433  &  12.0  &  10.3  &  11.1  \\
            6154.2170  &  11.0  &  2.102  &  -1.600  &  81.5  &  82.1  &  92.7  \\
            6160.7420  &  11.0  &  2.104  &  -1.260  &  105.4 &  107.6 &  111.4 \\
            5711.0910  &  12.0  &  4.346  &  -1.742  &  129.6 &  133.7 &  138.9 \\
            6318.7010  &  12.0  &  5.108  &  -2.020  &  68.0  &  71.3  &  77.3  \\
            6319.2280  &  12.0  &  5.108  &  -2.242  &  49.5  &  51.3  &  55.4  \\
            6319.4790  &  12.0  &  5.108  &  -2.719  &  25.0  &  26.2  &  29.5  \\
            6696.0110  &  13.0  &  3.143  &  -1.562  &  76.9  &  78.7  &  89.9  \\
            6698.6440  &  13.0  &  3.143  &  -1.830  &  54.0  &  55.0  &  65.4  \\
            5517.5420  &  14.0  &  5.082  &  -2.554  &  22.5  &  24.5  &  24.7  \\
            5645.6010  &  14.0  &  4.929  &  -2.120  &  56.9  &  54.7  &  58.7  \\
            5665.5710  &  14.0  &  4.920  &  -2.040  &  66.2  &  65.3  &  68.4  \\
            5684.4830  &  14.0  &  4.954  &  -1.700  &  78.7  &  74.0  &  73.9  \\
            5690.4270  &  14.0  &  4.929  &  -1.840  &  66.1  &  65.7  &  65.3  \\
            5701.1040  &  14.0  &  4.929  &  -2.080  &  53.7  &  56.2  &  55.6  \\
            5793.0610  &  14.0  &  4.929  &  -2.060  &  62.4  &  62.9  &  65.4  \\
            6125.0010  &  14.0  &  5.613  &  -1.580  &  44.8  &  43.7  &  43.7  \\
            6142.4760  &  14.0  &  5.619  &  -1.530  &  41.5  &  40.5  &  42.1  \\
            6145.0050  &  14.0  &  5.616  &  -1.450  &  45.6  &  44.5  &  46.1  \\
            6243.8070  &  14.0  &  5.616  &  -1.270  &  59.2  &  53.6  &  60.2  \\
            6244.4610  &  14.0  &  5.616  &  -1.340  &  52.8  &  56.1  &  64.2  \\
            6721.8170  &  14.0  &  5.861  &  -1.090  &  62.7  &  58.2  &  42.5  \\
            6741.5880  &  14.0  &  5.984  &  -1.750  &  24.0  &  23.0  &  -     \\
            6045.9620  &  16.0  &  7.867  &  -1.820  &  20.0  &  17.8  &  19.5  \\
            6743.5720  &  16.0  &  7.866  &  -0.920  &  -     &  10.5  &  -     \\
            5260.3920  &  20.0  &  2.521  &  -1.820  &  64.4  &  69.5  &  69.7  \\
            5512.9870  &  20.0  &  2.932  &  -0.300  &  109.6 &  114.3 &  116.2 \\
            5867.5590  &  20.0  &  2.932  &  -1.630  &  54.2  &  58.2  &  61.8  \\
            6161.2750  &  20.0  &  2.523  &  -1.293  &  114.0 &  113.1 &  -     \\
            6166.4290  &  20.0  &  2.521  &  -1.136  &  111.1 &  109.6 &  -     \\
            6455.5940  &  20.0  &  2.523  &  -1.320  &  107.1 &  102.9 &  111.2 \\
            4913.6340  &  22.0  &  1.873  &   0.161  &  92.4  &  95.8  &  101.1 \\
            5062.1150  &  22.0  &  2.160  &  -0.420  &  47.9  &  59.1  &  62.9  \\
            5113.4500  &  22.0  &  1.443  &  -0.820  &  77.3  &  79.2  &  84.9  \\
            5145.4740  &  22.0  &  1.460  &  -0.600  &  83.9  &  91.8  &  99.4  \\
            5219.7100  &  22.0  &  0.021  &  -2.292  &  99.7  &  105.4 &  115.1 \\
            5282.3990  &  22.0  &  1.053  &  -1.640  &  70.3  &  75.4  &  85.5  \\
            5295.7840  &  22.0  &  1.066  &  -1.680  &  65.4  &  70.2  &  77.5  \\
            5490.1530  &  22.0  &  1.460  &  -0.950  &  78.3  &  80.7  &  88.3  \\
            5648.5680  &  22.0  &  2.495  &  -0.350  &  44.2  &  44.4  &  54.3  \\
            5739.4730  &  22.0  &  2.249  &  -0.724  &  38.0  &  41.6  &  49.2  \\
            5922.1090  &  22.0  &  1.046  &  -1.450  &  89.1  &  91.5  &  105.4 \\
            5965.8060  &  22.0  &  1.879  &  -0.540  &  -     &  -     &  109.1 \\
            6091.1710  &  22.0  &  2.267  &  -0.430  &  61.9  &  63.3  &  72.8  \\
            6126.2100  &  22.0  &  1.066  &  -1.360  &  89.4  &  92.9  &  104.5 \\
            6258.0990  &  22.0  &  1.443  &  -0.340  &  106.3 &  117.6 &  -     \\
            6554.2140  &  22.0  &  1.443  &  -1.210  &  79.9  &  86.5  &  94.8  \\
            5211.5410  &  22.1  &  2.590  &  -1.456  &  50.0  &  49.1  &  45.0  \\
            5336.7950  &  22.1  &  1.582  &  -1.600  &  104.4 &  103.9 &  106.9 \\
            5418.7740  &  22.1  &  1.582  &  -2.080  &  80.2  &  83.8  &  81.6  \\
            5490.7010  &  22.1  &  1.566  &  -2.730  &  54.7  &  55.2  &  -     \\
            5670.8510  &  23.0  &  1.081  &  -0.578  &  98.4  &  104.4 &  119.1 \\
            4801.0440  &  24.0  &  3.121  &  -0.140  &  84.6  &  85.7  &  89.4  \\
            4936.3560  &  24.0  &  3.113  &  -0.280  &  78.8  &  82.6  &  87.4  \\
            4964.9460  &  24.0  &  0.941  &  -2.490  &  94.1  &  95.8  &  103.1 \\
            5214.1410  &  24.0  &  3.369  &  -0.753  &  37.7  &  33.8  &  39.5  \\
            5238.9700  &  24.0  &  2.709  &  -1.410  &  44.9  &  45.3  &  52.3  \\
            5272.0110  &  24.0  &  3.449  &  -0.482  &  51.7  &  58.2  &  60.8  \\
            5287.1810  &  24.0  &  3.438  &  -0.957  &  33.1  &  34.0  &  37.1  \\
            5300.7580  &  24.0  &  0.983  &  -2.120  &  117.7 &  116.8 &  -     \\
            5304.1830  &  24.0  &  3.463  &  -0.730  &  41.1  &  39.1  &  45.9  \\
            5318.7770  &  24.0  &  3.438  &  -0.720  &  42.5  &  41.9  &  -     \\
            5329.1530  &  24.0  &  2.913  &  -0.194  &  113.3 &  113.8 &  -     \\
            5628.6390  &  24.0  &  3.422  &  -0.790  &  41.9  &  44.0  &  46.4  \\
            5783.0650  &  24.0  &  3.323  &  -0.450  &  68.5  &  67.4  &  74.2  \\
            5783.8690  &  24.0  &  3.322  &  -0.335  &  94.8  &  95.6  &  102.2 \\
            6330.0830  &  24.0  &  0.941  &  -2.880  &  90.1  &  95.8  &  108.0 \\
            6661.0550  &  24.0  &  4.193  &  -0.365  &  31.3  &  29.9  &  36.7  \\
            5237.3200  &  24.1  &  4.073  &  -1.155  &  64.1  &  63.0  &  62.3  \\
            5246.7610  &  24.1  &  3.713  &  -2.436  &  27.4  &  26.2  &  -     \\
            5058.5100  &  26.0  &  3.642  &  -2.750  &  32.0  &  36.6  &  41.9  \\
            5180.0650  &  26.0  &  4.473  &  -1.120  &  84.7  &  82.5  &  87.6  \\
            5196.0670  &  26.0  &  4.256  &  -0.590  &  92.5  &  97.6  &  100.5 \\
            5197.9470  &  26.0  &  4.301  &  -1.480  &  55.8  &  59.4  &  58.3  \\
            5223.1930  &  26.0  &  3.635  &  -2.243  &  58.4  &  58.9  &  64.4  \\
            5243.7830  &  26.0  &  4.256  &  -0.930  &  85.0  &  86.8  &  89.6  \\
            5253.0310  &  26.0  &  2.279  &  -3.849  &  61.7  &  65.8  &  70.1  \\
            5288.5290  &  26.0  &  3.695  &  -1.550  &  94.4  &  94.7  &  97.7  \\
            5294.5560  &  26.0  &  3.640  &  -2.680  &  42.0  &  43.0  &  48.4  \\
            5295.3190  &  26.0  &  4.415  &  -1.530  &  54.4  &  53.5  &  55.3  \\
            5321.1170  &  26.0  &  4.435  &  -1.261  &  68.1  &  66.4  &  71.0  \\
            5326.1570  &  26.0  &  3.573  &  -2.210  &  77.1  &  77.5  &  84.5  \\
            5386.3370  &  26.0  &  4.154  &  -1.700  &  60.5  &  61.3  &  64.0  \\
            5401.2970  &  26.0  &  4.320  &  -1.720  &  60.7  &  59.4  &  70.3  \\
            5441.3440  &  26.0  &  4.312  &  -1.590  &  58.2  &  60.6  &  62.4  \\
            5460.8790  &  26.0  &  3.071  &  -3.530  &  30.6  &  34.3  &  38.4  \\
            5461.5500  &  26.0  &  4.445  &  -1.612  &  51.7  &  53.6  &  56.6  \\
            5464.2810  &  26.0  &  4.143  &  -1.582  &  70.4  &  72.7  &  71.4  \\
            5470.0970  &  26.0  &  4.445  &  -1.610  &  49.5  &  52.0  &  55.3  \\
            5494.4680  &  26.0  &  4.076  &  -1.960  &  59.3  &  61.8  &  64.9  \\
            5522.4500  &  26.0  &  4.209  &  -1.400  &  71.2  &  71.7  &  72.2  \\
            5525.5520  &  26.0  &  4.230  &  -1.184  &  87.0  &  86.7  &  84.2  \\
            5538.5200  &  26.0  &  4.217  &  -1.559  &  60.8  &  68.9  &  72.8  \\
            5539.2820  &  26.0  &  3.642  &  -2.610  &  43.8  &  50.0  &  57.0  \\
            5546.5140  &  26.0  &  4.371  &  -1.080  &  82.8  &  82.0  &  87.8  \\
            5552.6990  &  26.0  &  4.956  &  -1.800  &  18.2  &  16.8  &  18.0  \\
            5560.2140  &  26.0  &  4.435  &  -1.000  &  74.8  &  76.1  &  74.0  \\
            5587.5800  &  26.0  &  4.143  &  -1.650  &  67.6  &  69.7  &  71.9  \\
            5608.9850  &  26.0  &  4.209  &  -2.360  &  34.8  &  33.9  &  38.2  \\
            5611.3590  &  26.0  &  3.635  &  -2.960  &  34.0  &  34.8  &  36.3  \\
            5618.6370  &  26.0  &  4.209  &  -1.260  &  79.2  &  77.6  &  81.4  \\
            5619.6020  &  26.0  &  4.386  &  -1.480  &  66.5  &  63.7  &  71.0  \\
            5635.8270  &  26.0  &  4.256  &  -1.560  &  66.1  &  65.9  &  69.6  \\
            5636.7020  &  26.0  &  3.640  &  -2.530  &  51.8  &  55.7  &  60.0  \\
            5651.4720  &  26.0  &  4.473  &  -1.780  &  41.2  &  42.6  &  44.4  \\
            5652.3160  &  26.0  &  4.260  &  -1.760  &  53.7  &  54.4  &  59.6  \\
            5677.6910  &  26.0  &  4.103  &  -2.680  &  23.3  &  22.3  &  28.1  \\
            5679.0260  &  26.0  &  4.652  &  -0.680  &  81.2  &  79.7  &  81.1  \\
            5680.2340  &  26.0  &  4.186  &  -2.370  &  35.0  &  33.3  &  39.3  \\
            5705.4650  &  26.0  &  4.301  &  -1.455  &  69.5  &  72.1  &  74.1  \\
            5717.8300  &  26.0  &  4.284  &  -0.990  &  94.3  &  93.9  &  98.0  \\
            5731.7650  &  26.0  &  4.256  &  -1.060  &  89.7  &  90.8  &  92.3  \\
            5741.8480  &  26.0  &  4.256  &  -1.640  &  60.2  &  62.3  &  67.0  \\
            5742.9520  &  26.0  &  4.178  &  -2.320  &  35.8  &  36.1  &  41.9  \\
            5752.0330  &  26.0  &  4.548  &  -0.867  &  79.1  &  81.1  &  82.2  \\
            5760.3490  &  26.0  &  3.642  &  -2.450  &  60.4  &  60.2  &  58.4  \\
            5775.0790  &  26.0  &  4.220  &  -1.040  &  89.9  &  92.4  &  96.1  \\
            5853.1480  &  26.0  &  1.485  &  -5.170  &  53.6  &  56.6  &  63.1  \\
            5855.0770  &  26.0  &  4.608  &  -1.540  &  44.7  &  47.0  &  46.4  \\
            5856.0840  &  26.0  &  4.294  &  -1.550  &  65.3  &  65.7  &  67.2  \\
            5859.5860  &  26.0  &  4.548  &  -0.388  &  99.3  &  99.3  &  100.  \\
            5861.1040  &  26.0  &  4.283  &  -2.400  &  25.7  &  26.0  &  28.2  \\
            5881.2700  &  26.0  &  4.608  &  -1.770  &  40.0  &  40.0  &  42.6  \\
            5927.7860  &  26.0  &  4.652  &  -1.020  &  66.7  &  68.3  &  67.8  \\
            5929.6720  &  26.0  &  4.548  &  -1.170  &  68.5  &  69.4  &  70.3  \\
            5983.6800  &  26.0  &  4.548  &  -0.558  &  97.8  &  97.2  &  95.7  \\
            6007.9590  &  26.0  &  4.652  &  -0.620  &  87.7  &  87.1  &  89.2  \\
            6027.0510  &  26.0  &  4.076  &  -1.020  &  99.1  &  99.9  &  103.  \\
            6056.0000  &  26.0  &  4.733  &  -0.340  &  97.0  &  95.0  &  99.0  \\
            6079.0050  &  26.0  &  4.652  &  -0.960  &  71.4  &  75.3  &  73.6  \\
            6082.7020  &  26.0  &  2.223  &  -3.550  &  93.8  &  94.2  &  98.9  \\
            6093.6370  &  26.0  &  4.608  &  -1.340  &  57.9  &  57.2  &  59.3  \\
            6094.3630  &  26.0  &  4.652  &  -1.610  &  44.2  &  43.3  &  45.6  \\
            6096.6570  &  26.0  &  3.984  &  -1.810  &  74.1  &  75.3  &  75.1  \\
            6098.2460  &  26.0  &  4.558  &  -1.800  &  42.8  &  42.0  &  43.3  \\
            6120.2410  &  26.0  &  0.915  &  -5.930  &  51.1  &  54.2  &  64.5  \\
            6159.3600  &  26.0  &  4.608  &  -1.890  &  34.9  &  34.8  &  38.4  \\
            6165.3520  &  26.0  &  4.143  &  -1.430  &  81.2  &  78.2  &  83.0  \\
            6187.3840  &  26.0  &  2.831  &  -4.190  &  25.8  &  26.0  &  29.1  \\
            6187.9790  &  26.0  &  3.943  &  -1.660  &  87.2  &  85.0  &  87.4  \\
            6220.7740  &  26.0  &  3.881  &  -2.370  &  52.8  &  53.8  &  57.4  \\
            6226.7240  &  26.0  &  3.883  &  -2.110  &  61.8  &  64.0  &  64.8  \\
            6271.2760  &  26.0  &  3.332  &  -2.763  &  63.5  &  65.1  &  68.3  \\
            6311.4890  &  26.0  &  2.831  &  -3.150  &  74.8  &  73.1  &  80.0  \\
            6315.8000  &  26.0  &  4.076  &  -1.650  &  73.4  &  76.5  &  80.1  \\
            6330.8360  &  26.0  &  4.733  &  -1.170  &  59.7  &  58.9  &  60.5  \\
            6380.7400  &  26.0  &  4.186  &  -1.280  &  89.3  &  89.7  &  99.1  \\
            6392.5260  &  26.0  &  2.279  &  -3.956  &  67.2  &  68.0  &  76.0  \\
            6436.3790  &  26.0  &  4.186  &  -2.410  &  33.7  &  33.4  &  35.0  \\
            6496.4590  &  26.0  &  4.795  &  -0.520  &  90.1  &  91.0  &  94.0  \\
            6556.7730  &  26.0  &  4.795  &  -1.705  &  33.5  &  38.1  &  36.3  \\
            6581.1840  &  26.0  &  1.485  &  -4.750  &  91.6  &  91.2  &  -     \\
            6591.2930  &  26.0  &  4.593  &  -2.030  &  27.3  &  24.9  &  30.6  \\
            6608.0060  &  26.0  &  2.279  &  -3.990  &  68.3  &  70.2  &  73.8  \\
            6624.9980  &  26.0  &  1.011  &  -5.330  &  97.3  &  93.7  &  -     \\
            6627.5340  &  26.0  &  4.548  &  -1.500  &  58.0  &  58.8  &  64.0  \\
            6699.1240  &  26.0  &  4.593  &  -2.170  &  26.0  &  22.7  &  30.7  \\
            6703.5480  &  26.0  &  2.758  &  -3.010  &  90.2  &  90.3  &  96.0  \\
            6713.7220  &  26.0  &  4.795  &  -1.440  &  47.9  &  44.6  &  44.4  \\
            6725.3370  &  26.0  &  4.103  &  -2.220  &  44.4  &  45.1  &  49.5  \\
            6726.6500  &  26.0  &  4.607  &  -1.010  &  75.2  &  74.5  &  76.7  \\
            6733.1340  &  26.0  &  4.638  &  -1.440  &  54.5  &  54.4  &  56.0  \\
            6739.5030  &  26.0  &  1.557  &  -4.934  &  67.3  &  70.3  &  77.8  \\
            6745.9470  &  26.0  &  4.076  &  -2.730  &  23.3  &  24.4  &  27.6  \\
            6752.7050  &  26.0  &  4.638  &  -1.244  &  66.3  &  69.6  &  73.8  \\
            6786.8420  &  26.0  &  4.191  &  -1.920  &  58.7  &  57.6  &  59.0  \\
            5197.5830  &  26.1  &  3.230  &  -2.230  &  86.9  &  90.3  &  81.9  \\
            5234.6390  &  26.1  &  3.221  &  -2.180  &  92.2  &  88.3  &  86.1  \\
            5264.8120  &  26.1  &  3.230  &  -3.130  &  55.7  &  61.3  &  52.9  \\
            5325.5630  &  26.1  &  3.221  &  -3.210  &  54.7  &  54.4  &  53.4  \\
            5425.2550  &  26.1  &  3.199  &  -3.290  &  51.0  &  52.5  &  47.0  \\
            5991.3590  &  26.1  &  3.153  &  -3.590  &  47.8  &  47.4  &  47.2  \\
            6084.1030  &  26.1  &  3.199  &  -3.800  &  35.0  &  35.2  &  34.4  \\
            6149.2380  &  26.1  &  3.889  &  -2.750  &  43.9  &  43.7  &  42.5  \\
            6247.5480  &  26.1  &  3.892  &  -2.350  &  54.4  &  55.7  &  52.7  \\
            6369.4500  &  26.1  &  2.891  &  -4.180  &  32.2  &  32.9  &  28.9  \\
            6432.6670  &  26.1  &  2.891  &  -3.630  &  57.9  &  55.9  &  52.3  \\
            6456.3690  &  26.1  &  3.903  &  -2.080  &  70.1  &  70.4  &  63.1  \\
            4935.8440  &  28.0  &  3.941  &  -0.350  &  70.5  &  78.6  &  76.3  \\
            5010.9510  &  28.0  &  3.635  &  -0.870  &  67.9  &  73.4  &  71.6  \\
            5094.4190  &  28.0  &  3.833  &  -1.090  &  44.1  &  47.7  &  48.3  \\
            5102.9870  &  28.0  &  1.676  &  -2.810  &  95.5  &  97.5  &  100.8 \\
            5155.1330  &  28.0  &  3.898  &  -0.640  &  68.4  &  73.3  &  69.3  \\
            5197.1780  &  28.0  &  3.898  &  -1.240  &  53.1  &  57.0  &  60.1  \\
            5468.1130  &  28.0  &  3.847  &  -1.630  &  28.2  &  30.4  &  31.6  \\
            5578.7210  &  28.0  &  1.676  &  -2.790  &  113.9 &  115.4 &  -     \\
            5589.3590  &  28.0  &  3.898  &  -1.200  &  51.4  &  51.0  &  50.8  \\
            5643.0820  &  28.0  &  4.165  &  -1.237  &  32.5  &  33.4  &  35.4  \\
            5748.3480  &  28.0  &  1.676  &  -3.280  &  87.4  &  93.1  &  98.3  \\
            5760.8290  &  28.0  &  4.105  &  -0.780  &  64.4  &  64.6  &  64.6  \\
            5805.2150  &  28.0  &  4.167  &  -0.696  &  65.2  &  64.4  &  68.4  \\
            5996.7240  &  28.0  &  4.236  &  -1.040  &  43.1  &  43.5  &  45.7  \\
            6007.3070  &  28.0  &  1.676  &  -3.364  &  78.2  &  78.7  &  85.9  \\
            6086.2740  &  28.0  &  4.266  &  -0.470  &  68.2  &  69.6  &  71.0  \\
            6111.0630  &  28.0  &  4.088  &  -0.800  &  62.1  &  62.9  &  69.8  \\
            6128.9700  &  28.0  &  1.676  &  -3.330  &  83.9  &  82.7  &  95.2  \\
            6130.1290  &  28.0  &  4.266  &  -0.930  &  40.4  &  41.9  &  46.7  \\
            6176.8030  &  28.0  &  4.088  &  -0.210  &  92.1  &  93.4  &  97.2  \\
            6177.2390  &  28.0  &  1.826  &  -3.540  &  54.5  &  65.1  &  72.6  \\
            6204.5930  &  28.0  &  4.088  &  -1.110  &  49.8  &  51.5  &  57.2  \\
            6223.9690  &  28.0  &  4.105  &  -0.950  &  54.9  &  53.9  &  56.7  \\
            6322.1490  &  28.0  &  4.153  &  -1.220  &  36.3  &  39.1  &  37.9  \\
            6327.5900  &  28.0  &  1.676  &  -3.100  &  103.6 &  108.0 &  110.7 \\
            6378.2420  &  28.0  &  4.153  &  -0.830  &  60.2  &  60.0  &  65.3  \\
            6482.7810  &  28.0  &  1.935  &  -2.830  &  103.0 &  107.7 &  114.9 \\
            6586.2960  &  28.0  &  1.951  &  -2.800  &  102.6 &   01.0 &  111.7 \\
            6598.5780  &  28.0  &  4.236  &  -0.940  &  45.2  &  47.3  &  53.4  \\
            6635.0990  &  28.0  &  4.419  &  -0.770  &  50.1  &  50.6  &  49.7  \\
            6772.2930  &  28.0  &  3.658  &  -0.930  &  81.6  &  83.7  &  86.0  \\
            4722.1810  &  30.0  &  4.030  &  -0.338  &  77.3  &  83.4  &  82.8  \\
            4810.5470  &  30.0  &  4.078  &  -0.137  &  80.6  &  78.7  &  74.6  \\
            4883.6980  &  39.1  &  1.084  &   0.070  &  96.4  &  96.7  &  94.3  \\
            4900.1290  &  39.1  &  1.033  &  -0.090  &  94.4  &  95.9  &  96.1  \\
            5728.8940  &  39.1  &  1.839  &  -1.270  &  13.8  &  13.7  &  10.0  \\
\end{longtable}

\end{document}